\documentclass[aps,prd,twocolumn,nofootinbib,superscriptaddress,floatfix,longbibliography,10pt]{revtex4-2}

\usepackage{graphicx}
\usepackage{dcolumn}
\usepackage{footmisc}
\usepackage[normalem]{ulem}
\usepackage{bm}
\usepackage{xcolor}
\usepackage{multirow}
\usepackage{makecell}
\usepackage{physics}
\usepackage{float}
\usepackage[colorlinks=true,citecolor=blue,urlcolor=blue]{hyperref}
\usepackage{booktabs}
\usepackage{siunitx}

\begin{document}
\setlength{\abovedisplayskip}{5pt}
\setlength{\belowdisplayskip}{5pt}
\setlength{\abovedisplayshortskip}{5pt}
\setlength{\belowdisplayshortskip}{5pt}
\hyphenpenalty=1050


\title{Towards the Detection of Thermal Solar Neutrinos}

\author{Carlos A. Argüelles}
\email{carguelles@fas.harvard.edu}
\affiliation{Department of Physics \& Laboratory for Particle Physics and Cosmology,
Harvard University, Cambridge, MA 02138, USA}

\author{Christopher V. Cappiello}
\email{cappiello@wustl.edu}
\affiliation{Department of Physics and McDonnell Center for the Space Sciences, Washington University, St.~Louis, MO 63130, USA}

\author{P. S. Bhupal Dev}
\email{bdev@wustl.edu}
\affiliation{Department of Physics and McDonnell Center for the Space Sciences, Washington University, St.~Louis, MO 63130, USA}
\affiliation{PRISMA$^{++}$ Cluster of Excellence \& Mainz Institute for Theoretical Physics, 
Johannes Gutenberg-Universit\"{a}t Mainz, 55099 Mainz, Germany}

\author{Pablo Figueroa}
\email{pablofig@ific.uv.es}
\affiliation{Instituto de F\'isica Corpuscular (IFIC), CSIC-Universitat de Val\`encia, \\ Parc Científic UV, c/ Catedr\'atico Jos\'e Beltr\'an 2, E-46980 Paterna, Spain}

\author{Gonzalo Herrera}
\email{gonzaloh@mit.edu}
\affiliation{Department of Physics \& Laboratory for Particle Physics and Cosmology,
Harvard University, Cambridge, MA 02138, USA}
\affiliation{Department of Physics and Kavli Institute for Astrophysics and Space Research,
Massachusetts Institute of Technology, Cambridge, MA 02139, USA}
\affiliation{Center for Neutrino Physics, Department of Physics, Virginia Tech, Blacksburg, VA 24061, USA}

\author{Icarus Scoville}
\email{scovillekaitlyn07@student.fz.k12.mo.us}
\affiliation{Fort Zumwalt South High School, 8050 Mexico Rd, St.~Peters, MO 63376, USA}

\begin{abstract}
We show that $\sim$keV thermal solar neutrinos, arising from electroweak processes in the solar plasma,  are kinematically accessible to large-volume dark matter direct detection experiments via electron ionization signatures.
Using S2-only data from the XENONnT experiment, we place an upper limit on the thermal solar neutrino flux of $\eta \lesssim 1.2 \times 10^8$ times the standard model predicted value, while paired searches from XENONnT, LZ and PandaX give slightly weaker limits. The future XLZD experiment could improve these limits by orders of magnitude. 
While still far from a detection, this result establishes low-threshold direct detection experiments as a viable probe of the lowest-energy neutrino sources in astrophysics, with important implications for stellar physics and beyond. 
\end{abstract}

\maketitle

\noindent
{\textbf {\textit{Introduction.}}}---
Solar neutrinos provide a real-time, unobstructed probe into the Sun’s interior, serving as physical proof of stellar fusion while unlocking groundbreaking discoveries like neutrino flavor conversion (for a recent review, see \emph{e.g.}, Ref.~\cite{Xu:2022wcq}). The least energetic neutrinos detected to date are those produced in the nuclear $pp$-chain in the Sun with energies $E_{\nu} \gtrsim 200$\,keV, observed via elastic neutrino-electron scattering at Borexino~\cite{BOREXINO:2018ohr}, and at some dark matter (DM) direct detection experiments such as XENONnT~\cite{XENON:2022ltv}, PandaX-4T~\cite{PandaX:2024jjs} and LZ~\cite{LZ:2024zvo}.
Remarkably, no neutrinos with energies below $\mathord{\sim}100$\,keV have ever been detected  from any anthropogenic, terrestrial, or astrophysical source.

The last decade has seen DM direct detection experiments make important progress in lowering their energy thresholds in nuclear and electron recoil signatures~\cite{Essig:2022dfa}.
These lower thresholds have provided sensitivity to the sub-GeV DM mass regime and permitted the first detection of solar neutrinos via coherent elastic neutrino nucleus scattering (CE$\nu$NS) at PandaX-4T~\cite{PandaX:2024muv}, XENONnT~\cite{XENON:2024ijk, XENON:2026ydt}, and LZ~\cite{LZ:2025igz}---the so-called neutrino ``floor''~\cite{Billard:2013qya} or ``fog''~\cite{OHare:2021utq} for DM direct detection searches.

The striking progress in lowering the minimum energy thresholds at underground detectors, without compromising their large exposures, raises the question of whether astrophysical neutrinos of sub-$pp$ energies could leave detectable signatures, allowing us to push the low-energy neutrino astronomy frontier.

\begin{figure}[b!]
\centering
\includegraphics[width=0.98\linewidth]{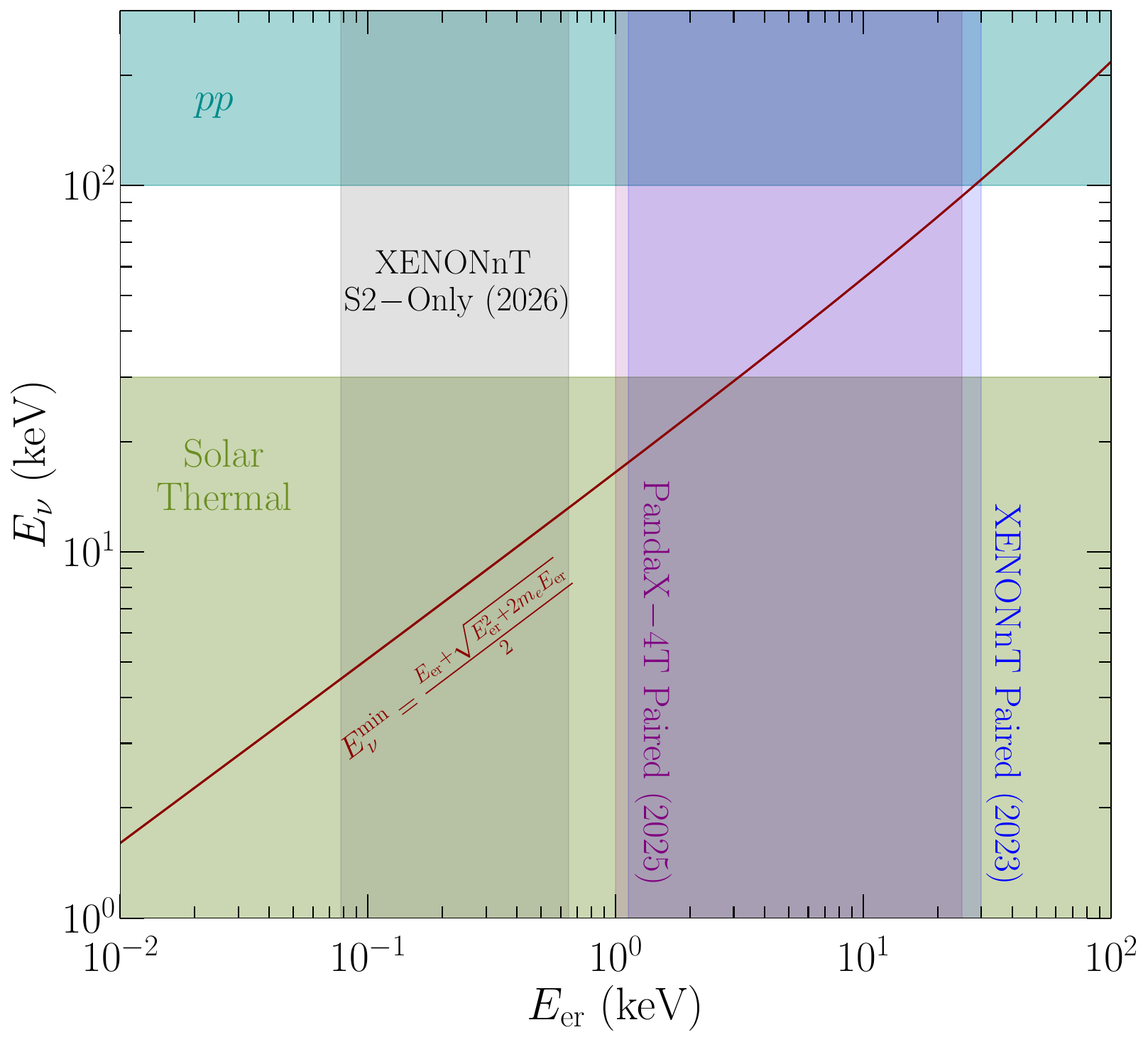}
\caption{Available kinematical regions of the XENONnT~\cite{XENON:2022ltv, XENON:2026qow} and PandaX-4T~\cite{PandaX:2024cic} experiments, together with the energy ranges of the $pp$ and thermal solar neutrinos.  In solid-red, the minimum neutrino energy required to induce a given recoil energy [cf.~Eq.~\eqref{enu_min}] is shown.}
\label{rate_pandax_xenon}
\end{figure}

Here, we make an  important observation that $\mathord{\sim}$\,keV thermal neutrinos from the Sun, arising from electroweak processes in the solar plasma~\cite{Adams:1963zzb,Haxton:2000xb,Vitagliano:2017odj}, can induce electron recoils falling within the region of interest (ROI) of direct detection searches, as illustrated in Fig.~\ref{rate_pandax_xenon}.
This observation allows us to derive the first experimental limits on the sub-100~keV astrophysical neutrino flux on Earth.  
We use thermal solar neutrino flux spectral predictions as a theoretical proxy and place limits on the overall normalization of this flux with current and future direct detection experiments.

\medskip
\noindent
{\textbf {\textit{Thermal solar neutrino flux.}}}---
Low-energy neutrinos can be produced in the solar plasma due to pair-production processes~\cite{Adams:1963zzb, Haxton:2000xb, Vitagliano:2017odj, Vitagliano:2019yzm}. 
The primary processes are photoproduction ($\gamma+e \rightarrow e+\nu\bar{\nu}$), bremsstrahlung (electron-ion or electron-electron scattering, $e+Ze\to Ze+e+\nu\bar{\nu}$), plasmon decay ($\gamma^{\star} \rightarrow \nu\bar{\nu}$), and $\nu\bar{\nu}$ emission in free-bound and bound-bound transitions associated with partially-ionized heavier elements. 
At sub-keV energies, plasmon decays dominate due to the large plasmon population in the solar core. 
The high-energy tail, from $\mathord{\sim} 1$ to $\mathord{\sim} 15$\,keV, is dominated by bremsstrahlung.
The thermal neutrino flux is expected to be larger than the low-energy tail of the nuclear $pp$ flux for neutrino energies below $\mathord{\sim} 3$\,keV.
The plasmon and bremsstrahlung channels are robust at the $10\%$ level, with the uncertainties dominated by the core temperature and electron/ion densities in the solar models~\cite{Vinyoles:2016djt}. See Supplemental Section~\ref{sec:thermalfluxvstemp} for further details on the thermal solar flux uncertainties.

The total thermal neutrino flux at Earth is $\Phi_{\rm th} \simeq 3.1 \times 10^{6}~\mathrm{cm}^{-2}\mathrm{s}^{-1}$, with an equal antineutrino flux~\cite{Vitagliano:2017odj}.  This is subdominant to the nuclear $pp$ flux ($\Phi_{pp} \mathord{\sim} 6 \times 10^{10}~\mathrm{cm}^{-2}\mathrm{s}^{-1}$), but extends to lower energies.  
The spectral shape peaks at $\mathord{\sim} 1$~keV for the bremsstrahlung contribution, while the plasmon component peaks at $\mathord{\sim} 0.3$~keV. 
The bound-bound and free-bound transitions associated with heavier elements (particularly iron) contribute in the few-keV range and carry information about the solar metallicity.

In this work, we parametrize the thermal neutrino flux by introducing a normalization factor $\eta$, such that the flux used in our calculations is $\eta$ times the Standard Solar Model prediction from Ref.~\cite{Vitagliano:2017odj}.  The theoretical expectation corresponds to $\eta = 1$.

The thermal neutrino flux represents a negligible energy-loss process for the Sun in the Standard Solar Model, with a total power of $3.5 \times 10^{25}$ erg/s. However, making the normalization factor $\eta$ too large would eventually imply a thermal neutrino luminosity inconsistent with the total energetics of the Sun. The most straightforward bound, though a somewhat conservative one, is set by requiring that the thermal neutrino luminosity not exceed the total luminosity of the Sun, $L_{\odot} = 3.85 \times 10^{33}$ erg/s (see \emph{e.g.}, the analogous argument regarding solar production of axions in Ref.~\cite{Caputo:2024oqc}). This argument constrains the maximum enhancement in the flux to be $\eta \lesssim 1.1 \times 10^8$.

\begin{figure}[t!]
\centering
\includegraphics[width=0.99\linewidth]{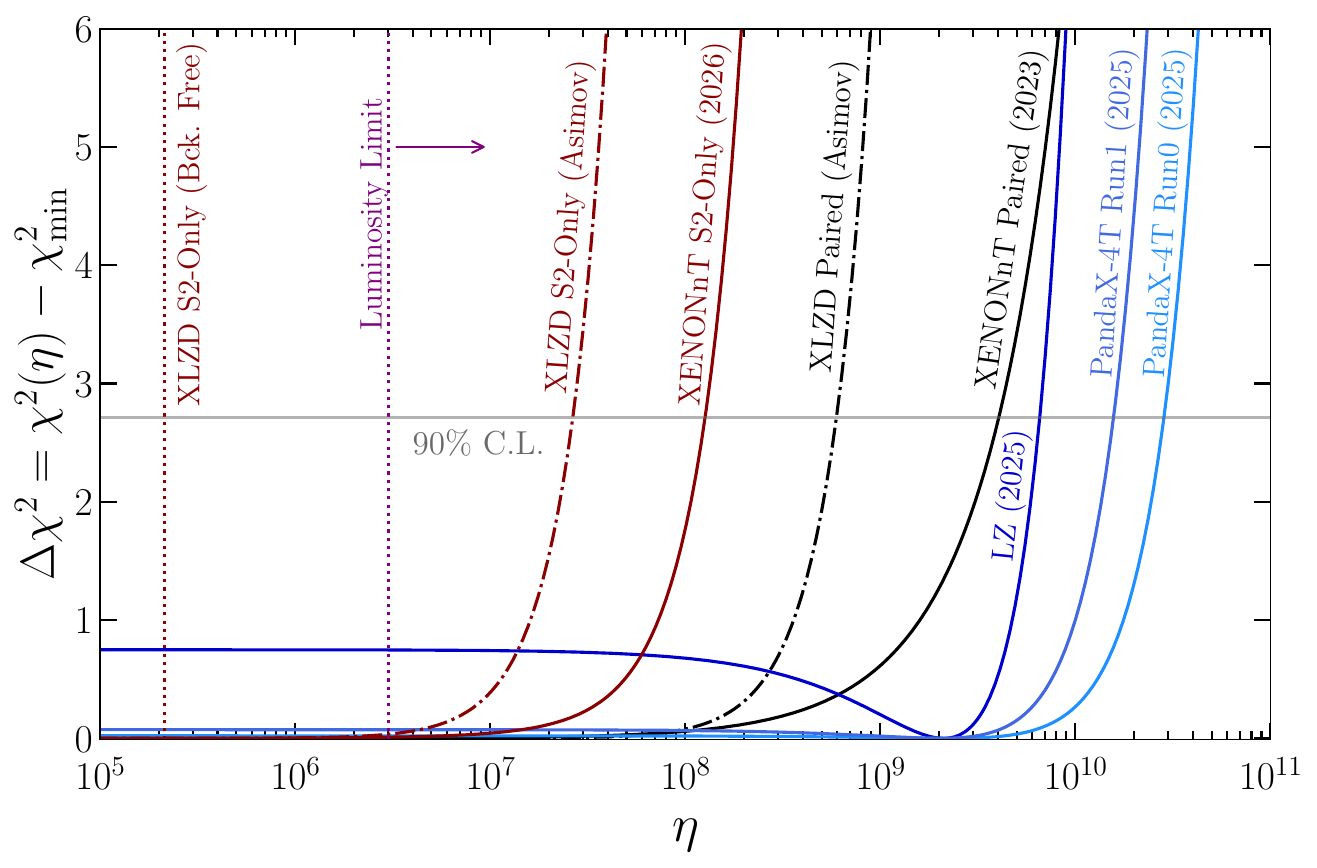}
\caption{$\Delta \chi^{2}$ as a function of the thermal solar neutrino flux normalization $\eta$.  For XENONnT, LZ and PandaX-4T, the data from all energy bins within the ROI were used.  The projection for XLZD assumes a $200$ ton-year exposure~\cite{Baudis:2024jnk}, but same threshold and ROI as XENONnT, and the corresponding rescaling for the background and signal events.  The horizontal dashed line indicates the $90\%$ C.L. threshold ($\Delta \chi^2 = 2.71$). We also show a background-free version of the XLZD projection by the vertical red dotted line. For comparison, the theoretical upper limit on $\eta$ from solar luminosity arguments is also shown.}
\label{fig:chi2_all}
\end{figure}

More stringent constraints on additional energy loss can be derived based on helioseismology and measured solar neutrino fluxes~\cite{Schlattl:1998fz,Raffelt:1999tx,Gondolo:2008dd,Vinyoles:2015aba}. These bounds restrict non-standard luminosity to be less than (0.03--0.1)~$L_{\odot}$, with some dependence on the particular energy-loss channel. The lower end of this range results in a limit $\eta \lesssim 3 \times 10^6$; see Fig.~\ref{fig:chi2_all}.

\medskip
\noindent
{\textbf {\textit{Elastic neutrino-electron scattering.}}}---
Neutrinos can elastically scatter off electrons in the atom.  The cross section can be calculated using the free-electron approximation, which accounts for the binding energy of electrons in the atom via a Heaviside step function~\cite{Essig:2018tss}.  In this approximation, the differential scattering cross section for neutrinos of flavor $\alpha = e, \mu, \tau$ on electrons is
\begin{align}
\frac{d\sigma_\alpha}{dE_{\rm er}} = \sum_{n,l} \theta(E_{\rm er} - |E^{nl}|) \frac{d\sigma^{0}_\alpha}{dE_{\rm er}} \,,
\label{eq:xsec_binding}
\end{align}
where the sum runs over all occupied atomic orbitals with binding energies $E^{nl}$ (see \emph{e.g.} Refs.~\cite{lbl, Essig:2017kqs} for tabulated values in xenon), and the cross section for neutrino scattering on free electrons ($\nu_\alpha + e \to \nu_\alpha + e)$ is given by~\cite{Vogel:1989iv}
\begin{align}
\frac{d\sigma^{0}_\alpha}{dE_{\rm er}} = \frac{2 G_F^2 m_e}{\pi} \bigg[g_{L\alpha}^2 \! + g_{R\alpha}^2 \bigg(\! 1-\frac{E_{\rm er}}{E_\nu} \!\bigg)^2 \! - g_{L\alpha} g_{R\alpha} \frac{m_e E_{\rm er}}{E_\nu^2}\bigg] .
\label{eq:xsec-sm-nu-e}
\end{align}
Here, $g_{Le} = (g_V + g_A)/2 + 1$, $g_{L\mu} = g_{L\tau} = (g_V + g_A)/2$, and $g_{R\alpha} = (g_V - g_A)/2$, with $g_V = -1/2 + 2 \sin^2\theta_w$, $g_A = -1/2$ and $\theta_w$ being the weak mixing angle.  For antineutrinos, $g_A \rightarrow -g_A$~\cite{Vogel:1989iv}.  The atomic binding effects yield a sizable suppression of the cross section at low recoil energies, particularly below $\mathord{\sim} 1$~keV where the inner-shell electrons of xenon become inaccessible. 
In such a regime, atomic corrections become relevant, and the validity of the Heaviside approximation employed here is only correct within $\mathord{\sim} 20-25 \%$~\cite{Chen:2016eab}.

The differential recoil rate is then
\begin{align}
\frac{dR}{dE_{\rm er}} = \epsilon\, N_T \sum_\alpha \!\! \int_{E_\nu^{\min}}^{E_\nu^{\max}} \!\! dE_\nu \frac{d\phi^\alpha}{dE_\nu} \frac{d\sigma_\alpha}{dE_{\rm er}},
\label{eq:dRdE-sm}
\end{align}
where $d\phi^\alpha/dE_\nu$ is the thermal neutrino flux for flavor $\alpha$, $\epsilon$ is the detection efficiency of the experiment, and $N_T$ is the number of target electrons. see Section~\ref{app:ionization} for details of the event rate calculation. The $pp$ contribution
is treated in the usual flavor basis, since it is produced as an almost pure $\nu_e$ source. For the thermal solar component, we use the mass-eigenstate fluxes $d\phi_i/dE_\nu$ at Earth, with $i=1,2,3$, rather than an effective flavor flux. 
This is the appropriate representation after propagation over the Sun--Earth baseline, where the thermal flux arrives as an incoherent mixture of mass eigenstates~\cite{Vitagliano:2019yzm}. 
We take the electron-flavor fractions to be  $P_{ei}=|U_{ei}|^2$, where $U$ is the PMNS mixing matrix, with
$P_{e1}=0.681$, $P_{e2}=0.297$, and $P_{e3}=0.022$, using the {\tt{NuFit6.0}} best-fit values of $U_{ei}$~\cite{Esteban:2024eli}, and project the arriving mass-eigenstate fluxes onto the weak-interaction channels relevant for elastic scattering on electrons through the PMNS matrix, as
\begin{align}
\frac{d\phi_{\nu_e}}{dE_\nu}
=
\sum_{i=1}^3 P_{ei}\frac{d\phi_i}{dE_\nu} \ , \ \ \ \
\frac{d\phi_{\nu_x}}{dE_\nu}
=
\sum_{i=1}^3 (1-P_{ei})\frac{d\phi_i}{dE_\nu},
\label{eq:thermal_mass_to_flavor}
\end{align}
where $\nu_x$ denotes the combined $\nu_\mu+\nu_\tau$ flux. Thermal emission proceeds through $\nu\bar\nu$-pair processes (plasmon decay,
pair annihilation, and bremsstrahlung), so we include an equal antineutrino
flux~\cite{Vitagliano:2019yzm}, folded with the corresponding
antineutrino--electron cross sections, which differ from the neutrino ones through the exchange of the $(g_V+g_A)^2$ and $(g_V-g_A)^2$ kinematic terms in Eq.~\eqref{eq:xsec-sm-nu-e}.

The minimum neutrino energy required to induce a given electron recoil energy $E_{\rm er}$ is
\begin{align}\label{enu_min}
E_\nu^{\min} = \frac{E_{\rm er} + \sqrt{2 m_e E_{\rm er} + E_{\rm er}^2}}{2} \,.
\end{align}
For keV-scale thermal neutrinos scattering off xenon electrons, the typical electron recoil energies are in the sub-keV to few-keV range, which falls within the ROI of recent ionization-only analyses. In Fig.~\ref{rate_pandax_xenon}, we show the neutrino energy ranges of the $pp$ and thermal solar neutrinos, together with the ROI for different electron-recoil analyses from XENON and PandaX collaborations.  The S2-only data release of XENONnT~\cite{XENON:2026qow} shows that the thermal solar neutrinos are already kinematically accessible for a tonne-year exposure DM experiment.  For comparison, we also show the ROI of the XENONnT and PandaX-4T paired analyses~\cite{XENON:2022ltv, PandaX:2024cic} (LZ~\cite{LZ:2025igz} has a similar ROI), for which the high-energy tail of the thermal solar neutrinos is already accessible.

\medskip
\noindent
{\textbf {\textit{A limit on the thermal neutrino flux from direct detection experiments.}}}---The total neutrino-electron scattering event rate in each electron recoil energy bin as a function of  the normalization of the thermal neutrino flux $\eta$, is given by
\begin{equation}\label{Ntot}
s_{i}(\eta) = \eta \ \mathcal{E} \int_{E_{R}^{\textrm{min},i}}^{E_{R}^{\textrm{max},i}}  \frac{dR}{d E_{\textrm{er}}} dE_{\textrm{er}},
\end{equation}
where $\mathcal{E}$ is the exposure, and $E_{R}^{\textrm{min,i}}$ and $E_{R}^{\textrm{max,i}}$ are the edges of each bin. For each of the considered experiments, we calculate the bin size $\Delta E_{\textrm{er}}^{(i)}$, and compute the total number of events for the corresponding energy bin. For nuclear recoil data, we compute the electron recoil-equivalent energy as described in Section\ref{app:ionization}. The observed and expected background events for each experiment are listed in Section\ref{app:complementary}. 

From the computation of the neutrino-electron scattering event rate, we turn to deriving upper limits on the normalization of the thermal solar neutrino flux.  We define a Poisson log-likelihood function as
\begin{equation}\label{eq:binned_likelihood}
\ln \mathcal{L}(\eta) = \sum_i \left[ n_i \ln(s_i + b_i) - (s_i + b_i) - \ln(n_i!) \right],
\end{equation}
where $n_{i}$ is the number of observed events in each bin, $b_{i}$ is the number of expected background events (reported by each collaboration), and $s_{i}=s_{i}(\eta)$ is the predicted signal from Eq.~\eqref{Ntot}.  We then define a test statistic based on the profile log-likelihood ratio
\begin{equation}
\chi^2(\eta) \equiv -2 \ln \mathcal{L}(\eta) \,,
\end{equation}
and find a $90\%$ C.L. upper bound on $\eta$ from
\begin{equation}\label{delta_chi2}
\Delta \chi^2 (\eta) = \chi^{2}(\eta) - \chi^{2}_{\rm min} \leq 2.71\,.
\end{equation}
When considering under-fluctuations ($n_{i} < b_{i}$), we follow the Feldman-Cousins prescription~\cite{Feldman_1998} to properly handle the boundary effects.  Under this framework, we derive upper limits on the thermal neutrino flux normalization from XENONnT, PandaX-4T (Run0 and Run1), LZ, and projections for DARWIN/XLZD~\cite{Baudis:2024jnk}, as reported in Table~\ref{tab:upper_limits}. The corresponding $\Delta \chi^2$ profiles are shown in Fig.~\ref{fig:chi2_all}.

\begin{table}[t!]
\centering
\sisetup{table-format=1.3}
\begin{tabular}{l S[table-format=1.3] S[table-format=1.1] S[table-format=1.2] S[table-format=1.1] S[table-format=1.1] S[table-format=2.0] S[table-format=2.0]}
\toprule
\multirow{2}{*}{Parameter}
& \multicolumn{2}{c}{XLZD}
& \multicolumn{2}{c}{XENONnT}
& {\multirow{2}{*}{LZ}}
& \multicolumn{2}{c}{PandaX-4T} \\
\cmidrule(lr){2-3} \cmidrule(lr){4-5} \cmidrule(lr){7-8}
& {S2-only} & {Paired} & {S2-only} & {Paired} & & {Run 1} & {Run 0} \\
\midrule
\multicolumn{1}{c}{$\eta\ (10^{9})$} & 0.027 & 0.6 & 0.12 & 4.1 & 6.5 & 16 & 28 \\
\bottomrule
\end{tabular}
\caption{Upper limits (projections) on the thermal neutrino flux normalization $\eta$ in units of $10^{9}$ from XENONnT, LZ, and PandaX-4T (XLZD). The strongest current limit comes from the XENONnT S2-only analysis.}
\label{tab:upper_limits}
\end{table}

The strongest current limit, $\eta \lesssim 1.2 \times 10^{8}$, comes from the XENONnT S2-only analysis, which benefits from its lower energy threshold reaching into the sub-keV regime where the thermal neutrino flux is largest.  The XENONnT paired analysis yields a weaker constraint of $\eta \lesssim 4.1 \times 10^{9}$ due to its higher threshold, thus sampling a lower density of thermal solar neutrinos, and also due to a lower exposure. Similarly,  LZ gives a limit of $\eta\lesssim 6.5 \times 10^{9}$, while PandaX-4T Run1 (Run0) provides a limit of $\eta \lesssim 1.6~(2.8) \times 10^{10}$. All these limits are comparably worse to XENONnT S2-only analysis due to a combination of somewhat higher thresholds, larger backgrounds, and lower exposures.  The projected XLZD S2-only sensitivity can reach $\eta \lesssim 2.7 \times 10^{7}$, an improvement of almost an order of magnitude over the current XENONnT limit, driven by its substantially larger exposure. We further perform a background-free XLZD analysis, finding that this could further extend the sensitivity reach to  $\eta \lesssim 2.1 \times 10^{5}$. In Section\ref{app:complementary}, we provide further details on the treatment of the LZ, PandaX-4T and the XENONnT paired data, for the derivation of the complementary bounds.

\begin{figure}[t!]
\centering
\includegraphics[width=0.99\linewidth]{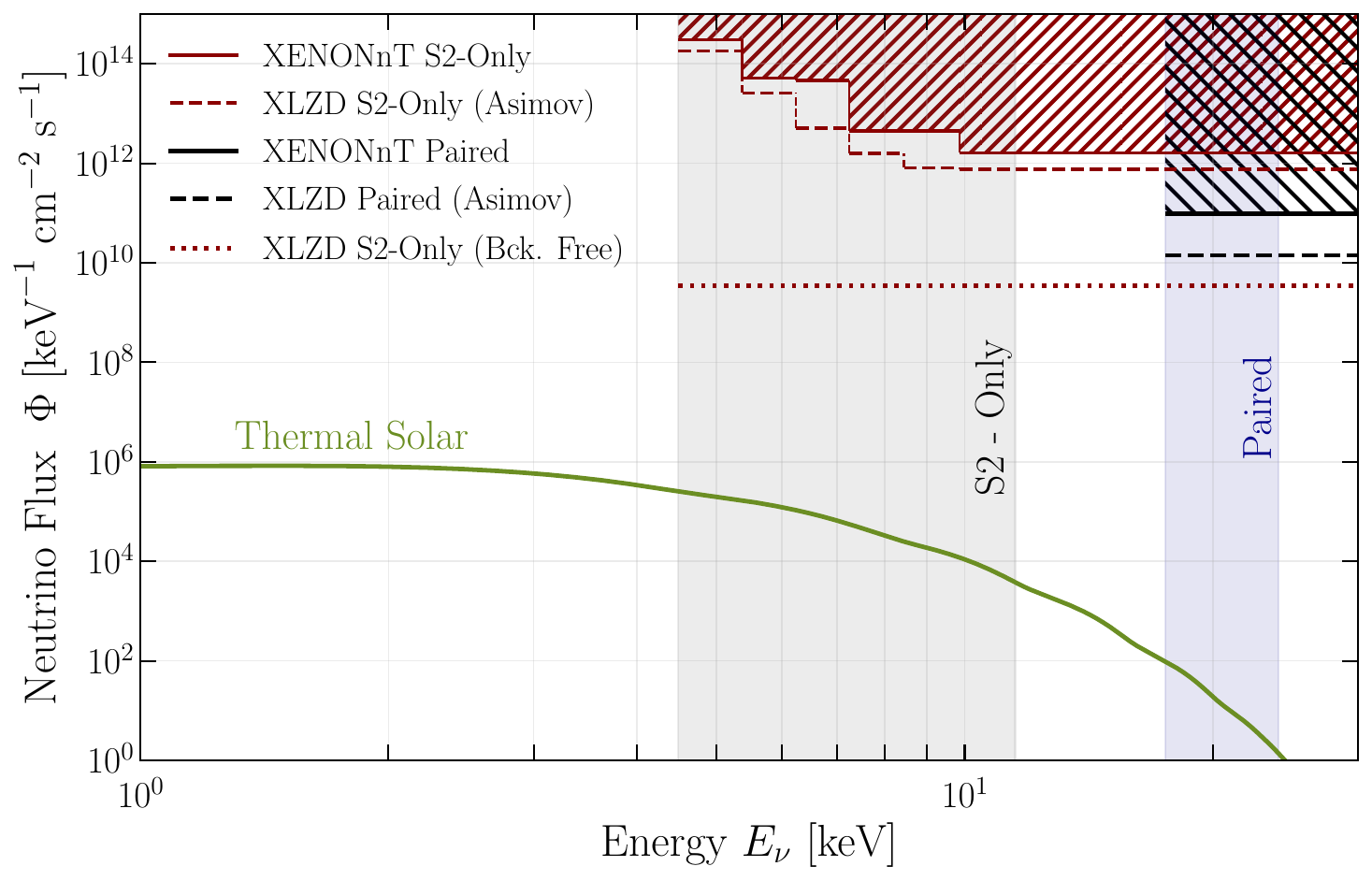}
\caption{Flux-equivalent $90 \%$ C.L. upper bounds (hatched) and projections (dashed) on the keV astrophysical neutrino flux derived in this work, confronted with the expected thermal solar neutrino flux. 
}
\label{fig:limit}
\end{figure}

 In Fig.~\ref{fig:limit}, we show the standard model  thermal neutrino flux spectrum prediction $(\eta = 1)$, along with the bin-wise limits arising from both S2-only and paired XENONnT data~\cite{XENON:2026qow, XENON:2022ltv}, and projections for the XLZD experiment; for details on the derivation of these limits, see Section~\ref{app:complementary}. We multiply the bin-wise (in electron recoil energy $E_{\rm er}$) normalization limit by the thermal solar-neutrino flux averaged over the neutrino-energy interval associated with that bin, spanning from $E_{\nu}^{\min, i}(E_{\rm er})$ up to the endpoint of the thermal neutrino flux, as the recoil rate in a given energy bin at XENONnT receives contributions from all neutrinos with energies above the threshold energy. The horizontal bands therefore represent flux-equivalent bin sensitivities integrated over the energy range above a certain threshold energy, not strict direct differential limits on $\Phi_{\textrm{th}}(E_\nu)$. The dashed XLZD lines correspond to the projected sensitivities displayed in Fig.~\ref{fig:chi2_all}, obtained from background-only Asimov data sets.

While these limits are still far from the theoretical prediction ($\eta = 1$), they represent the first experimental constraints on the keV astrophysical neutrino flux and demonstrate that direct detection experiments can extend the low-energy neutrino frontier well below the MeV scale. 

We note that the dominant irreducible background for a thermal solar neutrino flux search arises from elastic $pp$ neutrino-electron scattering (already considered in the experimental analysis). We considered a possible astrophysical background from cosmic-ray-boosted relic neutrino background component, but found that this is orders of magnitude below the thermal solar flux expectations; see Section~\ref{sec:CRboosted}.

\medskip
\noindent
{\textbf {\textit{Neutrino fog for light DM-electron scatterings.}}}---The thermal solar neutrino flux constitutes an irreducible background---a ``neutrino fog''---for light DM searches via electron recoils.  While the neutrino fog from nuclear $pp$-chain neutrinos has been extensively studied for both nuclear recoil~\cite{OHare:2021utq} and electron recoil~\cite{Carew:2023qrj, Essig:2018tss} channels, including the Migdal effect~\cite{Herrera:2023xun}, the contribution from thermal solar neutrinos has not been previously considered.

Since the thermal neutrino spectrum peaks at $\mathord{\sim} 1$~keV and extends down to sub-keV energies, it induces electron recoils in the same energy range as very light DM candidates with masses $m_{\rm DM} \lesssim 1$\,MeV scattering off electrons.  The neutrino-induced electron recoil rate from thermal neutrinos scales linearly with exposure, making it an irreducible background.

The neutrino fog can be quantified by comparing the DM-electron scattering rate with the neutrino-electron scattering rate.  For a DM candidate of mass $m_{\rm DM}$ interacting with electrons via a reference cross section $\bar{\sigma}_e$, the fog is reached when the DM signal becomes comparable to the systematic uncertainty on the neutrino background.

For neutrino energies below $\mathord{\sim} 10$~keV, the thermal solar neutrino flux dominates over the $pp$ neutrino tail, which may extend the neutrino fog to lower DM masses than previously estimated.  A detailed calculation of this extended fog is beyond the scope of this work but will be presented elsewhere.  Qualitatively, the thermal neutrino fog is expected to become relevant for DM masses $m_{\rm DM} \lesssim 100$~keV~\cite{Hertel:2018aal}, where the recoil energy spectrum from DM-electron scattering overlaps with the thermal neutrino-induced recoil spectrum. This might become relevant for future ultra-low-threshold DM direct detection experiments~\cite{Essig:2022dfa}.

\begin{figure}[t!]
    \centering
    \includegraphics[width=1.0\linewidth]{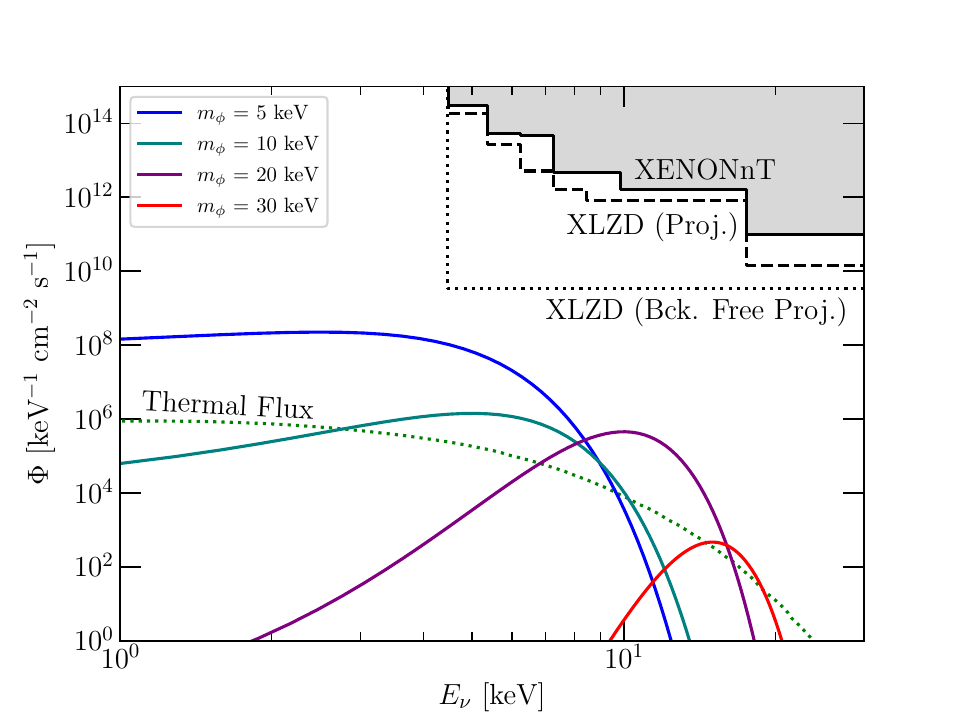}
    \caption{Neutrino flux induced by the decay of light scalars (of different masses) emitted from the Sun, assuming their coupling to electrons saturated by the red giant cooling constraint~\cite{Yamamoto:2023zlu}.  
    For comparison, we also show the SM thermal solar flux (dotted curve), the current XENONnT upper limit (shaded region) and the projected XLZD limits from Fig.~\ref{fig:limit}. }
    \label{fig:majoron1}
\end{figure}

\medskip
\noindent
{\textbf {\textit{New physics contributions.}}}---The sensitivity of direct detection experiments to keV-scale neutrino interactions opens the door to probing various new physics scenarios. 
As an illustration, we show in Fig.~\ref{fig:majoron1} the neutrino flux from the decay of a keV-scale scalar that is produced inside the Sun~\cite{Yamamoto:2023zlu}; see Section~\ref{E1} for details. We find that although this new physics component cannot yet be constrained by the direct detection experiments considered here, it can yield a keV astrophysical neutrino flux well above the expected thermal solar neutrino flux, thus further motivating the search for keV neutrinos.

Decaying keV sterile neutrino DM~\cite{Drewes:2016upu} provides another source of keV neutrino flux (see Section~\ref{E2}),  but the decay $J$-factor needed to exceed the thermal solar flux requires a spiked DM  profile~\cite{Gondolo:1999ef} and directional sensitivity of the direct detection experiments to the galactic center. 

The detection of thermal solar neutrinos would also open other avenues to test new physics in the neutrino sector. For instance, active-sterile neutrino oscillations with baseline $L=1$ AU and energy $E\sim$ keV has $L/E\sim 10^{15}~{\rm eV}^{-2}$ that would cover new regions of quasi-Dirac neutrino parameter space between $pp$ neutrinos and galactic neutrinos~\cite{MacDonald:2025jbm}. The low-energy of these neutrinos may also enable to probe nonstandard neutrino interactions enhanced at low momentum transfers, such as in models with light mediators~\cite{Schwemberger:2022fjl, Blanco-Mas:2024ale, DeRomeri:2024iaw,AtzoriCorona:2025gyz}, or with large neutrino magnetic moments~\cite{Babu:2020ivd,Giunti:2024gec,Pages:2025odc,Fajfer:2026gtz}.

\medskip
\noindent
{\textbf {\textit{Outlook.}}}---
The eventual detection of thermal solar neutrinos would provide the first direct evidence of thermal pair-production processes in the solar core and offer a unique probe of the solar interior complementary to helioseismology and nuclear $pp$-chain neutrino measurements. 
It would also provide direct evidence of stellar cooling via neutrino emission, so far only indirectly inferred through stellar evolution arguments, and could offer sensitivity to the plasmon decay rate in a stellar plasma.

While the limits obtained in this work remain far from a detection of the thermal solar neutrino flux, future kilo-tonne-year xenon exposures~\cite{Baudis:2024jnk} or novel ultra-low-threshold detection technologies~\cite{Essig:2022dfa, Ruzi:2023cvp} may bring this goal within reach. We would like to remind that the first direct detection experiment to ever run, about 40 years ago~\cite{Ahlen:1987mn}, set a constraint on WIMPs that were about $\mathord{\sim}$ 11 orders of magnitude worse than current limits~\cite{Baudis:2025yva}. 
Improvements of such caliber in astro-particle physics are definitely possible.

The scattering signatures studied here are complementary to probes of thermal solar neutrinos through neutrino capture, which are explored in a concurrent companion study \cite{FerrariInPreparation}.

\medskip
\noindent
{\textbf {\textit{Acknowledgments.}}}---
We thank Karthik Ramanathan and Volodymyr Takhistov for useful discussions on low-threshold direct detection experiments. CAA are supported by the Faculty of Arts and Sciences of Harvard University, Canadian Institute for Advanced Research (CIFAR), the National Science Foundation (NSF), the John Templeton Foundation, the Research Corporation for Science Advancement, and the David \& Lucile Packard Foundation. CVC was generously supported by Washington University in St. Louis through the Edwin Thompson Jaynes
Postdoctoral Fellowship. The work of BD was partly supported by the U.S. Department of Energy under grant No.~DE-SC0017987 and by a Humboldt Fellowship from the Alexander von Humboldt Foundation. The work of PF was supported by the Spanish National Grant PID2022-
137268NA-C55. The work of GH was supported by the Neutrino Theory Network Fellowship with contract number 726844, and by the U.S. Department of Energy under award number DE-SC0020262.
PF and GH also acknowledge support by the Munich Institute for Astro-, Particle and BioPhysics (MIAPbP), which is funded by the Deutsche Forschungsgemeinschaft (DFG, German Research Foundation) under Germany's Excellence Strategy -- EXC-2094 -- 390783311.

\bibliography{References}

\onecolumngrid
\medskip
\hrule
\begin{center}
{\it \large Supplemental Material}
\end{center}
\setcounter{equation}{0}
\setcounter{figure}{0}
\setcounter{table}{0}
\setcounter{section}{0}
\makeatletter
\renewcommand{\theequation}{S\arabic{equation}}
\renewcommand{\thefigure}{S\arabic{figure}}
\renewcommand{\thetable}{S\arabic{table}}
\renewcommand{\thesection}{S\arabic{section}}

\section{Uncertainties on the Thermal Neutrino Flux}
\label{sec:thermalfluxvstemp}

In principle, there is some uncertainty in the thermal solar neutrino flux stemming from uncertainties in the Sun's temperature profile. While these uncertainties are expected to be at the 10\% level~\cite{Vitagliano:2017odj}, we confirm this by computing the variation in the neutrino flux between different solar models. Following Ref.~\cite{Vitagliano:2017odj}, we use the Saclay model~\cite{Turck-Chieze:2001aug} as our benchmark model, and compare with the result of Ref.~\cite{Serenelli:2009yc} using the abundances of Ref.~\cite{Asplund:2009fu}. 

\begin{figure}[h!]
    \centering
    \includegraphics[width=0.75\linewidth]{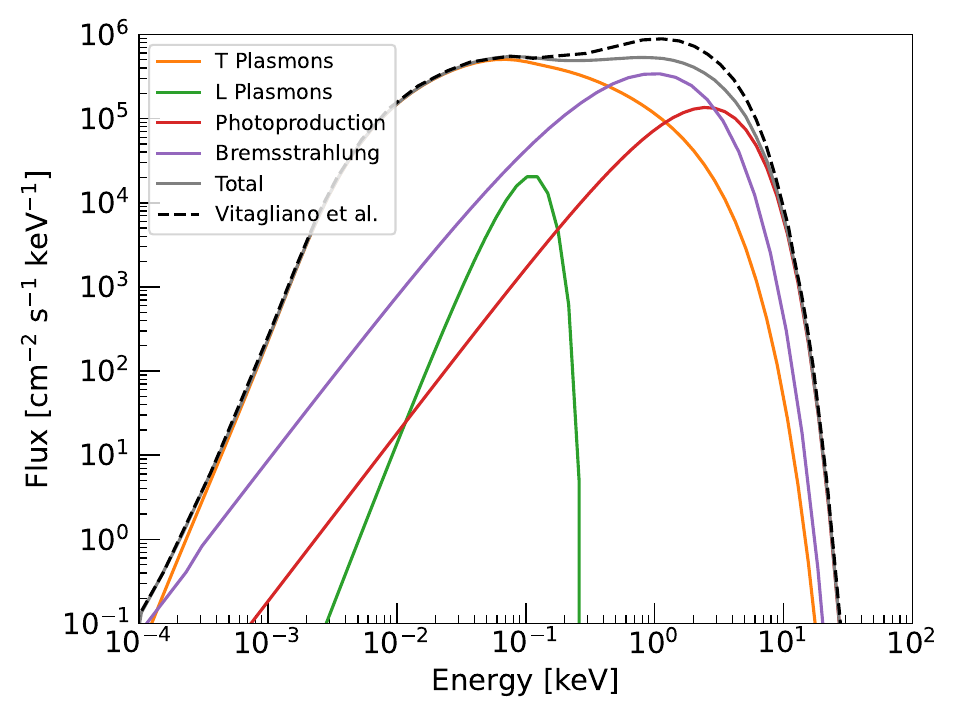}
    \caption{Different components contributing to the thermal solar neutrino flux, along with the sum of these components (gray). We compare our calculation to the result presented in Refs.~\cite{Vitagliano:2017odj,Vitagliano:2019yzm}. The difference between our results is primarily due to the lack of free-bound and bound-bound processes in our calculation.}
    \label{fig:components}
\end{figure}

Figure~\ref{fig:components} shows the contributions to the thermal neutrino flux from plasmon decay, photo-production, and bremsstrahlung. We only include free-free processes in the bremsstrahlung rate, which results in a slight discrepancy with Ref.~\cite{Vitagliano:2017odj} around 1 keV, but this discrepancy shrinks at larger energy where detectors such as XENONnT have the best sensitivity.

\begin{figure}
    \centering
    \includegraphics[width=0.5\linewidth]{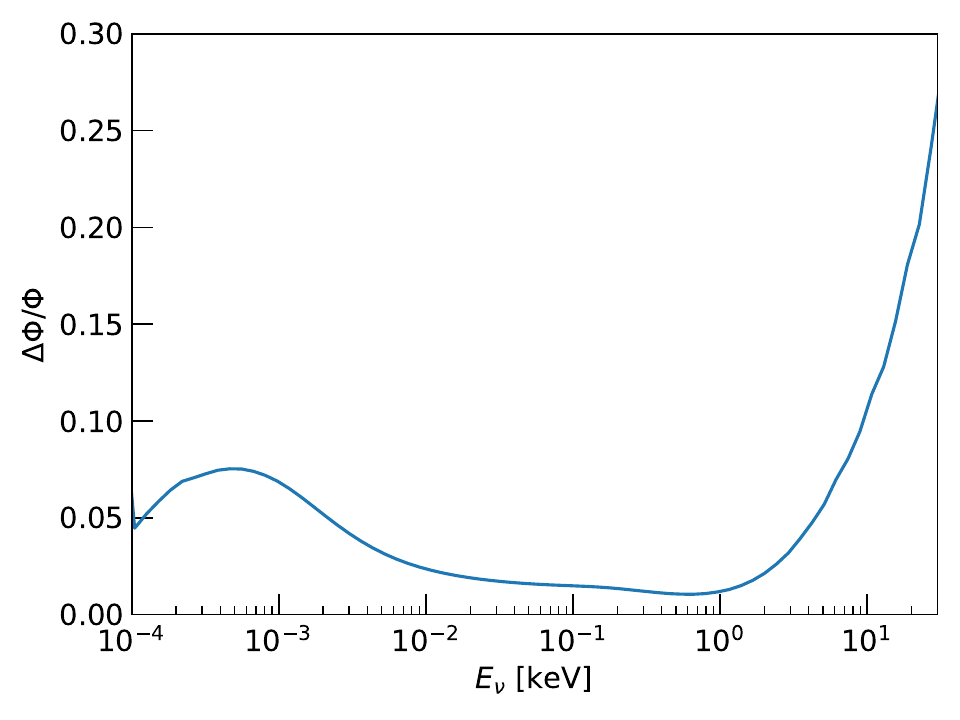}
    \caption{Fractional difference in the total thermal neutrino flux from the Sun for two different models for the solar temperature profile: that from the Saclay solar model~\cite{Turck-Chieze:2001aug} and that of Ref.~\cite{Serenelli:2009yc} using the abundances of Ref.~\cite{Asplund:2009fu}.}
    \label{fig:fractionaldifference}
\end{figure}

In Fig.~\ref{fig:fractionaldifference}, we show the fractional variation in the thermal neutrino flux between the two models we consider. For most of the energy range, the variation is less than 10\%. Only at the highest energies does the variation reach 20\% or more. This shows that the general uncertainty on the thermal solar neutrino flux due to uncertainties in the solar temperature profile is small.
\section{XENONnT ionization-only data from 2026 \label{app:ionization}}

The conversion from nuclear recoil to electron recoil-equivalent energy follows the procedure described in Ref.~\cite{Blanco-Mas:2024ale}, and the energy thresholds for PandaX-4T~\cite{PandaX:2024muv} and XENON1T~\cite{XENON:2019gfn} are taken from the respective experimental publications. 
In January 2026, the XENONnT collaboration released an ionization-only (S2-only) analysis searching for light DM~\cite{XENON:2026qow}.  This analysis achieves a significantly lower energy threshold than the standard paired (S1+S2) analysis by utilizing only the ionization (S2) signal, which extends the sensitivity to electron recoil energies below $\sim 1$\,keV$_{\rm ee}$.  This low threshold is crucial for our analysis, as it brings the peak of the thermal solar neutrino spectrum within the detectable range.

\begin{figure}[h]
\centering
    \includegraphics[width=0.49\linewidth]{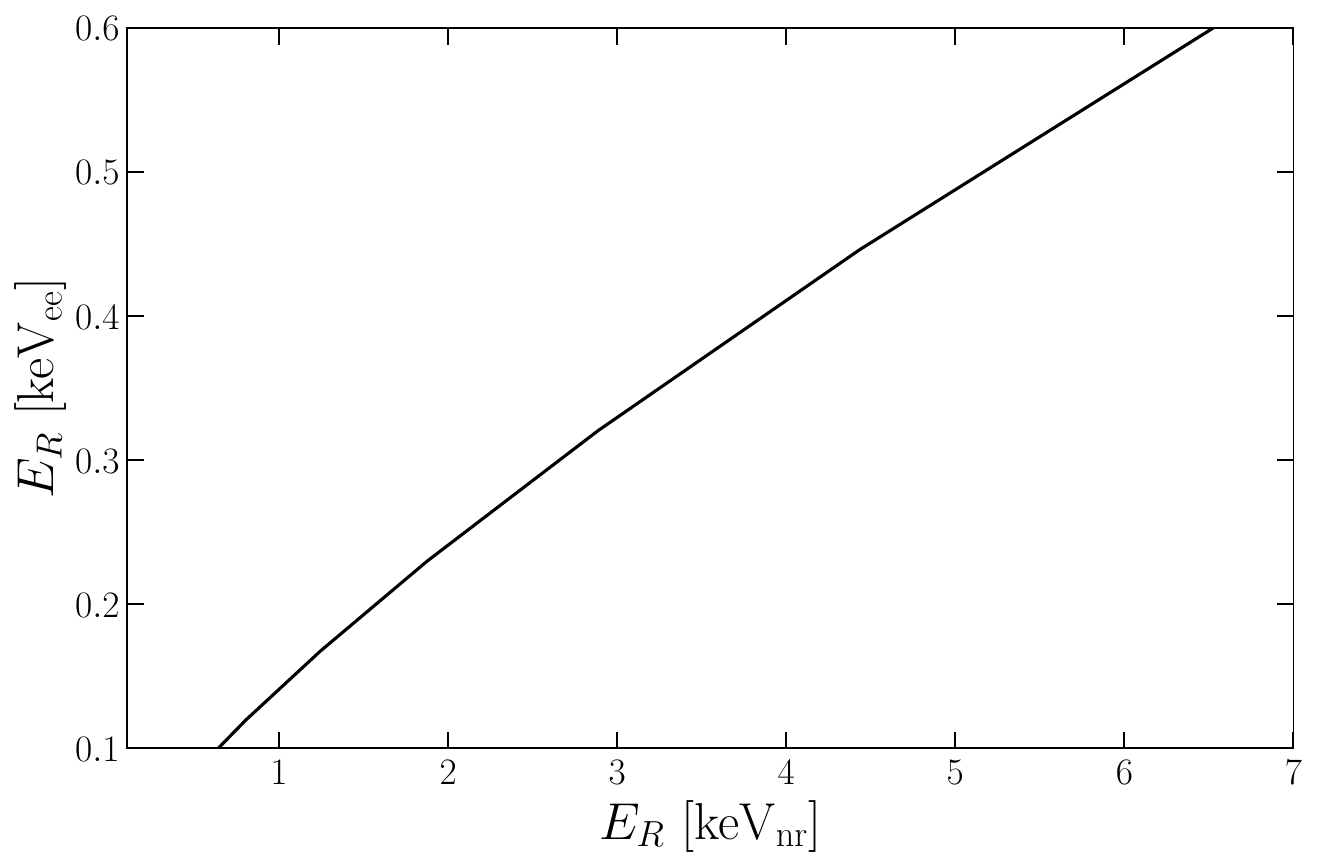}
    \hfill
    \includegraphics[width=0.49\linewidth]{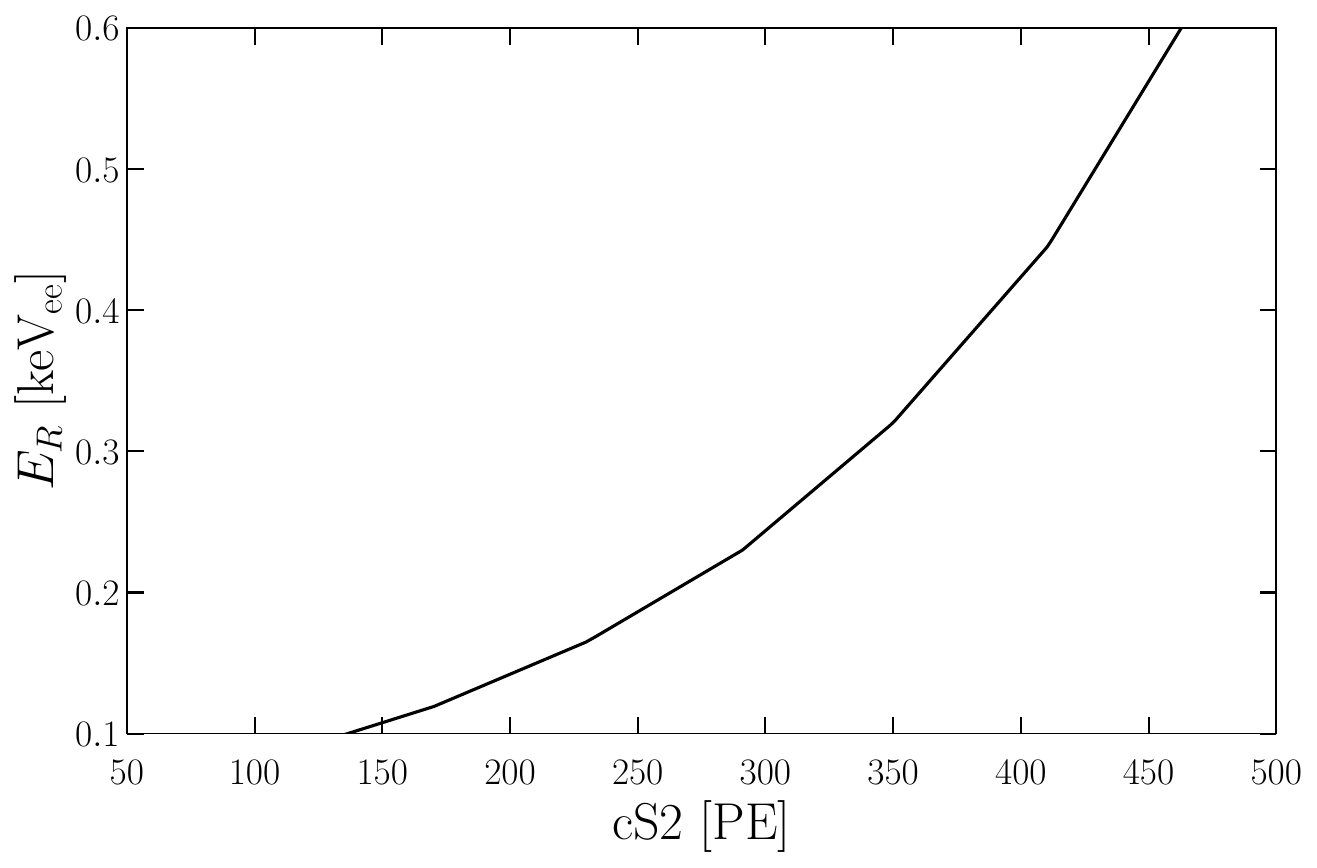}
    \caption{Electron recoil energy in electron equivalent $E_{R} \ [\textrm{keV}_{\textrm{ee}}]$, as a function of the nuclear recoil energy in nuclear equivalent $E_{R} \ [\textrm{keV}_{\textrm{nr}}]$ (left) and photo-electrons $c\textrm{S}2 \ [\textrm{PE}]$ (right)~\cite{XENON:2026qow}.}
\label{fig:quenching}
\end{figure}

The conversion between nuclear recoil energy $E_{\rm nr}$ and electron recoil-equivalent energy $E_{\rm ee}$ requires knowledge of the quenching factor, which describes the fraction of the deposited energy that produces ionization.  In Fig.~\ref{fig:quenching} we show the relationship between these energy scales as extracted from the XENONnT data.

\begin{figure}[h]
\centering
    \includegraphics[width=0.49\linewidth]{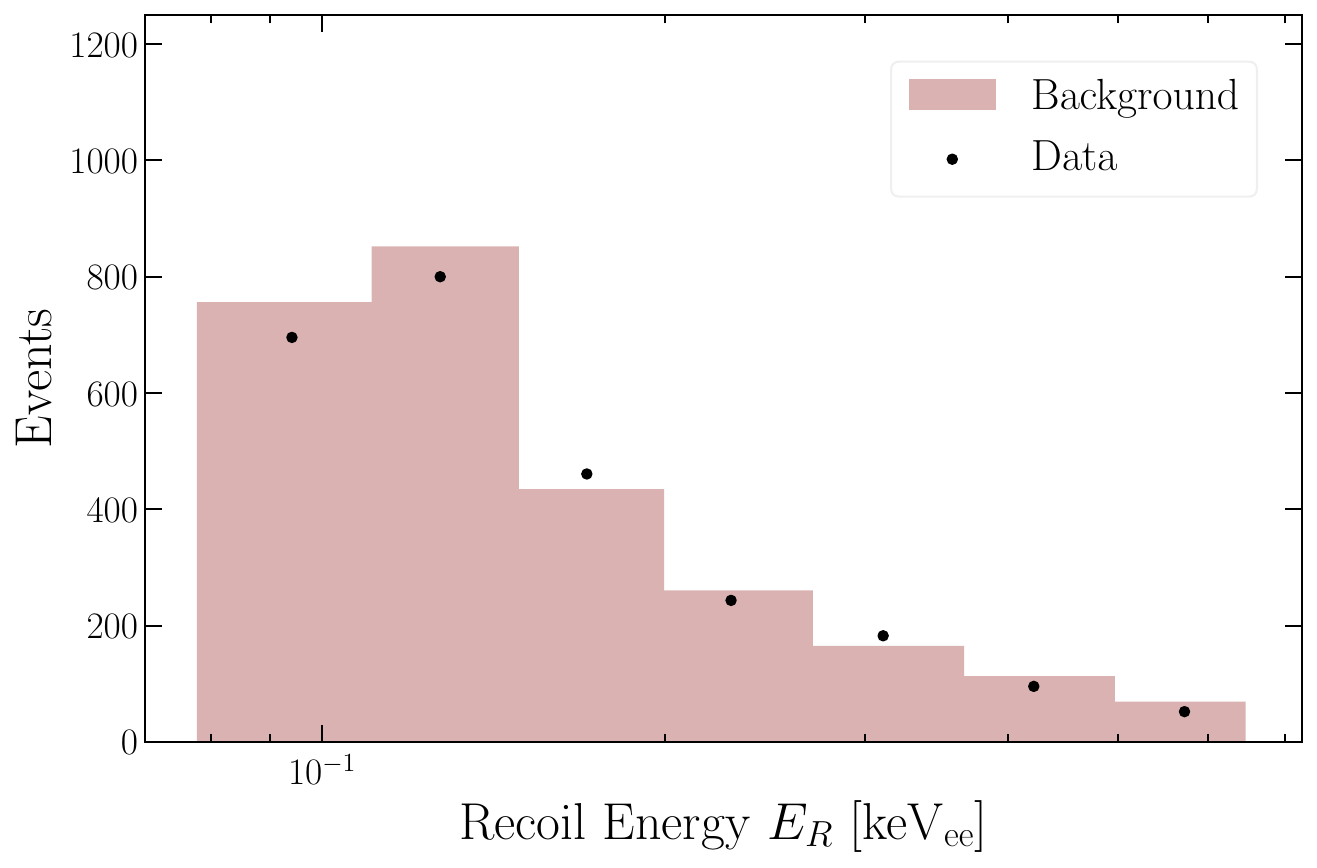}
    \hfill
    \includegraphics[width=0.49\linewidth]{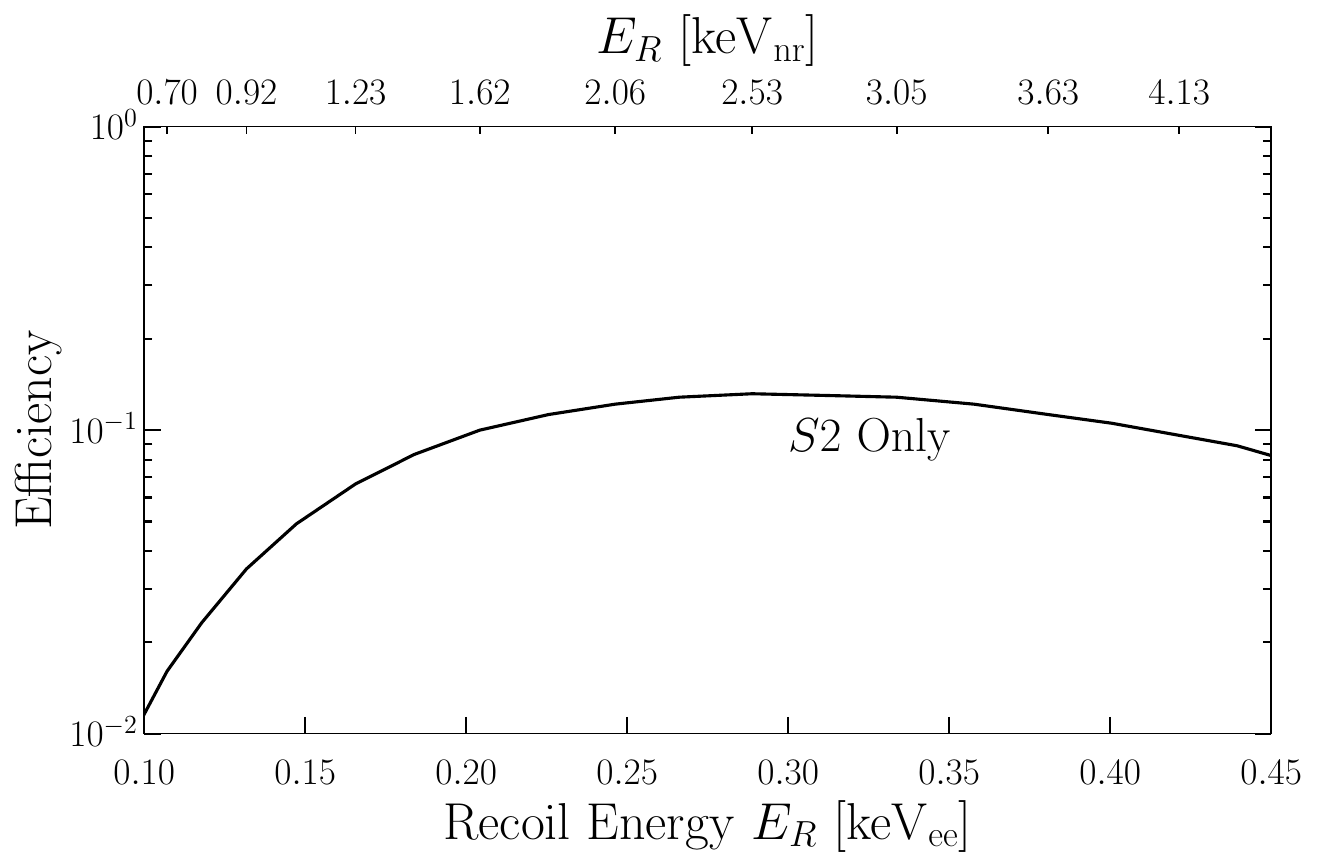}
    \caption{\textit{Left panel:} Total number of events as a function of the recoil energy for the XENONnT S2-Only (2026) science ROI \cite{XENON:2026qow}. \textit{Right panel:} Corresponding efficiency function as a function of the electron recoil energy in electron equivalent $E_{R} \ [\textrm{keV}_{\textrm{ee}}]$. For comparison in the upper $x$-axis the nuclear recoil energy is displayed $E_{R} \ [\textrm{keV}_{\textrm{nr}}]$.}
\label{fig:efficiency}
\end{figure}

The detection efficiency as a function of electron recoil energy is a critical input to our rate calculation.  The efficiency function accounts for trigger efficiency, data quality cuts, and analysis selection criteria.  At the lowest energies ($E_{\rm er} \lesssim 0.5$\,keV$_{\rm ee}$), the efficiency drops rapidly, while it plateaus near unity above $\sim 2$\,keV$_{\rm ee}$.  The efficiency curves are shown in Fig.~\ref{fig:efficiency}.

For the S2-only analysis, we compute the thermal neutrino signal rate by convolving the differential rate in Eq.~\eqref{eq:dRdE-sm} with the efficiency function and integrating over the energy bins defined by the experimental analysis.  The background model and observed event counts are taken directly from Ref.~\cite{XENON:2026qow}.

\begin{figure}[t!]
\centering
\includegraphics[width=0.5\linewidth]{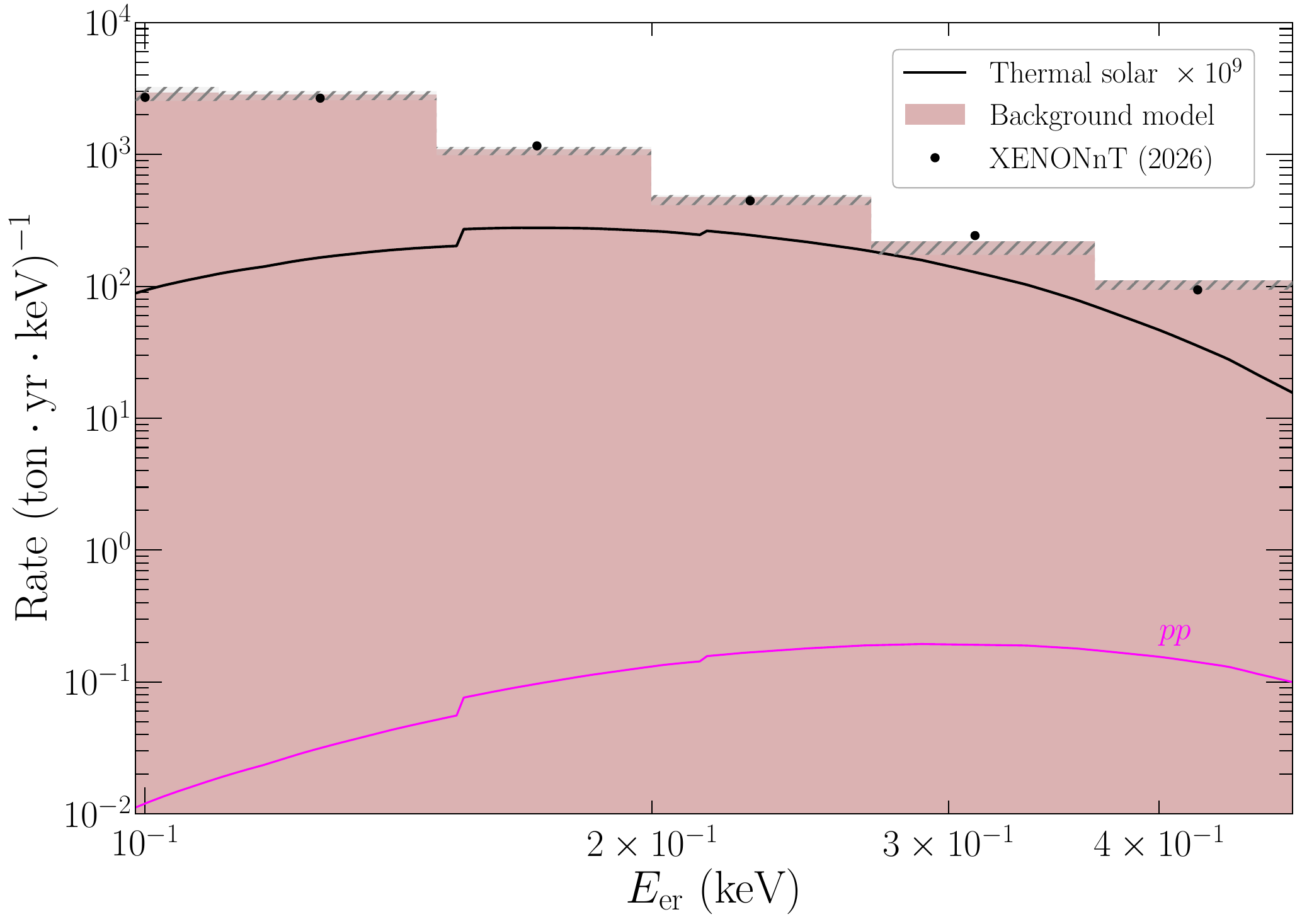}
\caption{Ionization event rate in the ROI of the S2-only data in XENONnT~\cite{XENON:2026qow} as a function of the electron recoil energy from both $pp$ (solid magenta) and thermal solar neutrinos (solid black).  For thermal neutrinos, the $\eta$ parameter was set to $10^{9}$, for visualization purposes.}
\label{fig:rate_event}
\end{figure}

 We show the resulting event rate in Fig.~\ref{fig:rate_event}, where the thermal neutrino signal (scaled by $\eta = 10^9$ for visibility) is compared with the $pp$ neutrino contribution and the experimental background model of the XENONnT S2-only analysis~\cite{XENON:2026qow}.

\section{Complementary Limits from XENONnT, LZ and PandaX-4T}\label{app:complementary}

In this section, we present the complementary limits on the thermal neutrino flux normalization from a previous XENONnT paired (S1+S2) analysis~\cite{XENON:2022ltv} ,LZ \cite{LZ:2025igz} and PandaX-4T Run0 and Run1~\cite{PandaX:2024cic}. 

The XENONnT paired (S1+S2) analysis~\cite{XENON:2022ltv} has a higher energy threshold ($E_{\rm er} \gtrsim 1$\,keV$_{\rm ee}$) compared to their S2-only analysis~\cite{XENON:2026qow}, but benefits from better background rejection due to the dual-signal requirement.  The PandaX-4T~\cite{PandaX:2024cic} and LZ~\cite{LZ:2025igz}  analyses provide independent constraints with different systematic uncertainties. In Table.~\ref{tab:bck_obs_events}, we summarize the observed and background number of events for the considered experiments.  For the case of XLZD~\cite{Baudis:2024jnk, XLZD:2024nsu}, we assume a 200-tonne-year exposure and a linear scaling of the current background and observed events of the XENONnT-Paired data.

\begin{table}[h!]
\centering
\begin{tabular}{
    l
    S[table-format=4.0]
    S[table-format=1.0]
    S[table-format=2.0]
    S[table-format=2.0]
    S[table-format=2.0]
}
\toprule
{\multirow{2}{*}{Events}}
& \multicolumn{2}{c}{XENONnT}
& {\multirow{2}{*}{LZ}}
& \multicolumn{2}{c}{PandaX-4T} \\
\cmidrule(lr){2-3}
\cmidrule(lr){5-6}
& {S2-only} & {Paired}
& & {Run 0} & {Run 1} \\
\midrule
Obs. & 2530 & 4 & 66 & 46 & 27 \\
Bkg. & 2652 & 8 & 59 & 45 & 26 \\
\bottomrule
\end{tabular}
\caption{Central values of the expected background (Bkg.) and observed (Obs.) number of events for the considered experiments, as reported in~\cite{XENON:2026qow, XENON:2024ijk, LZ:2025igz, PandaX:2024muv}. The XLZD Asimov entries are obtained by exposure-rescaling the corresponding XENONnT background counts and setting $\mathrm{Obs}=\mathrm{Bkg}$.}
\label{tab:bck_obs_events}
\end{table}

\begin{figure}[h]
\centering
\includegraphics[scale=0.165]{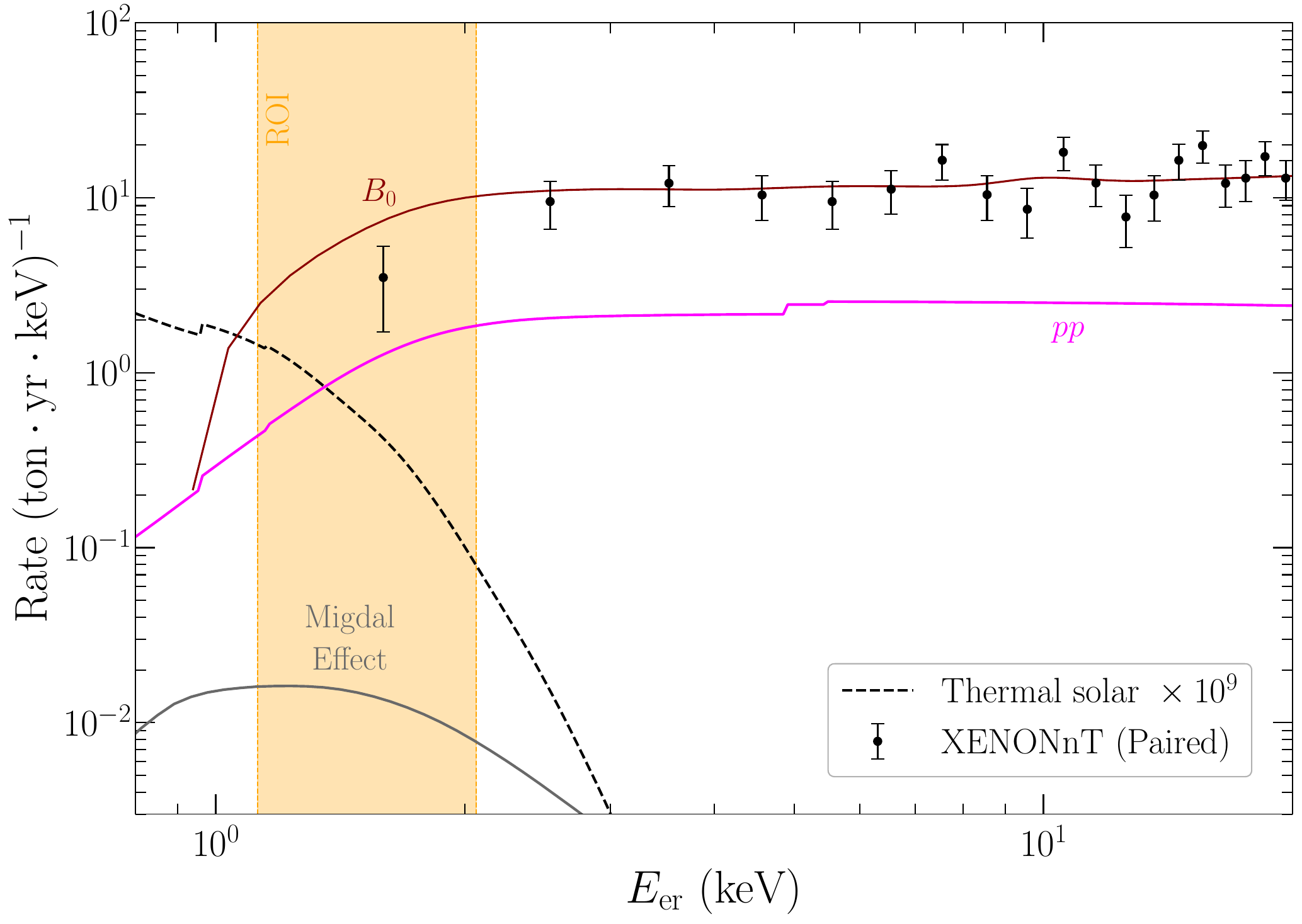}
\includegraphics[scale=0.165]{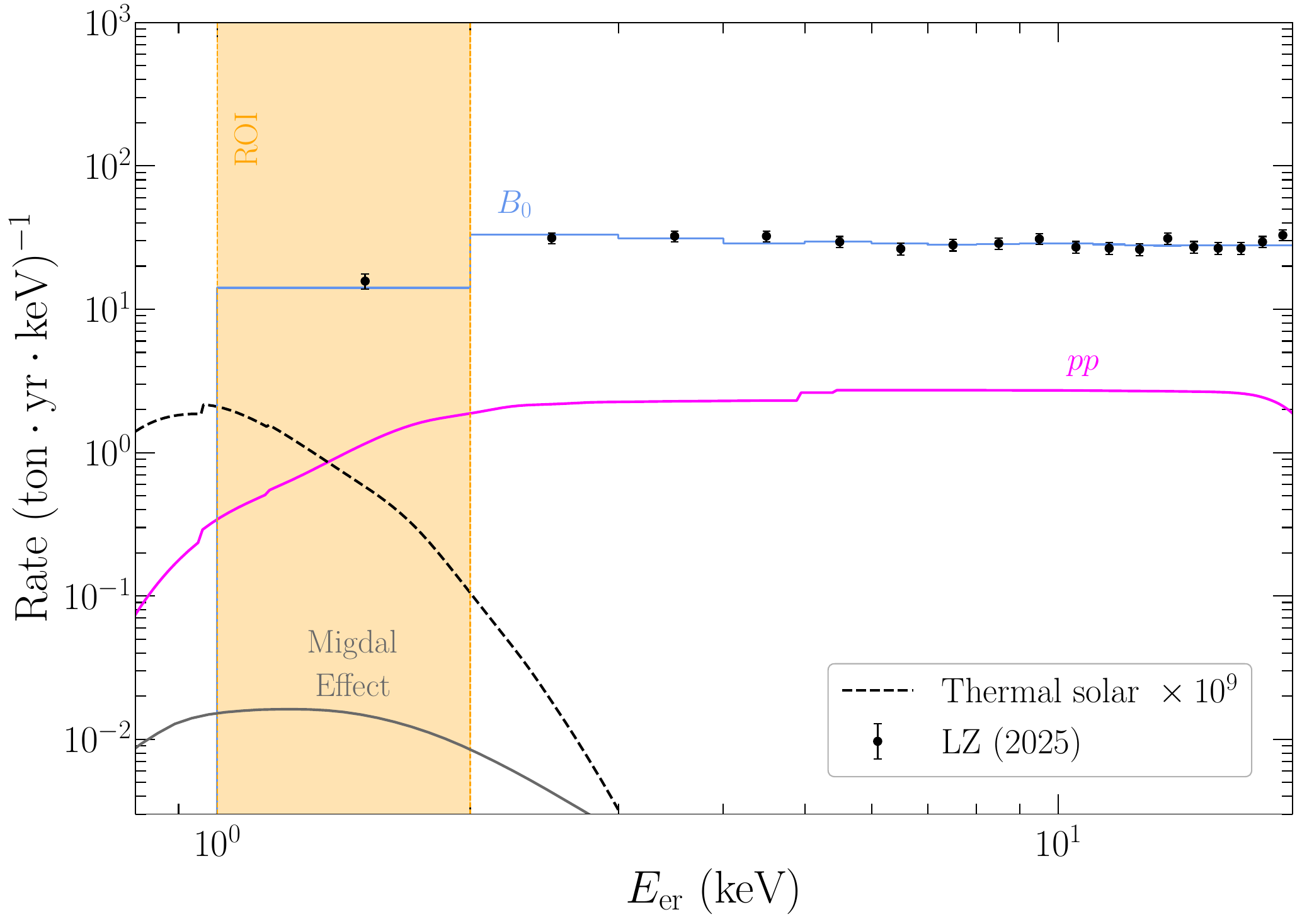}
\includegraphics[scale=0.165]{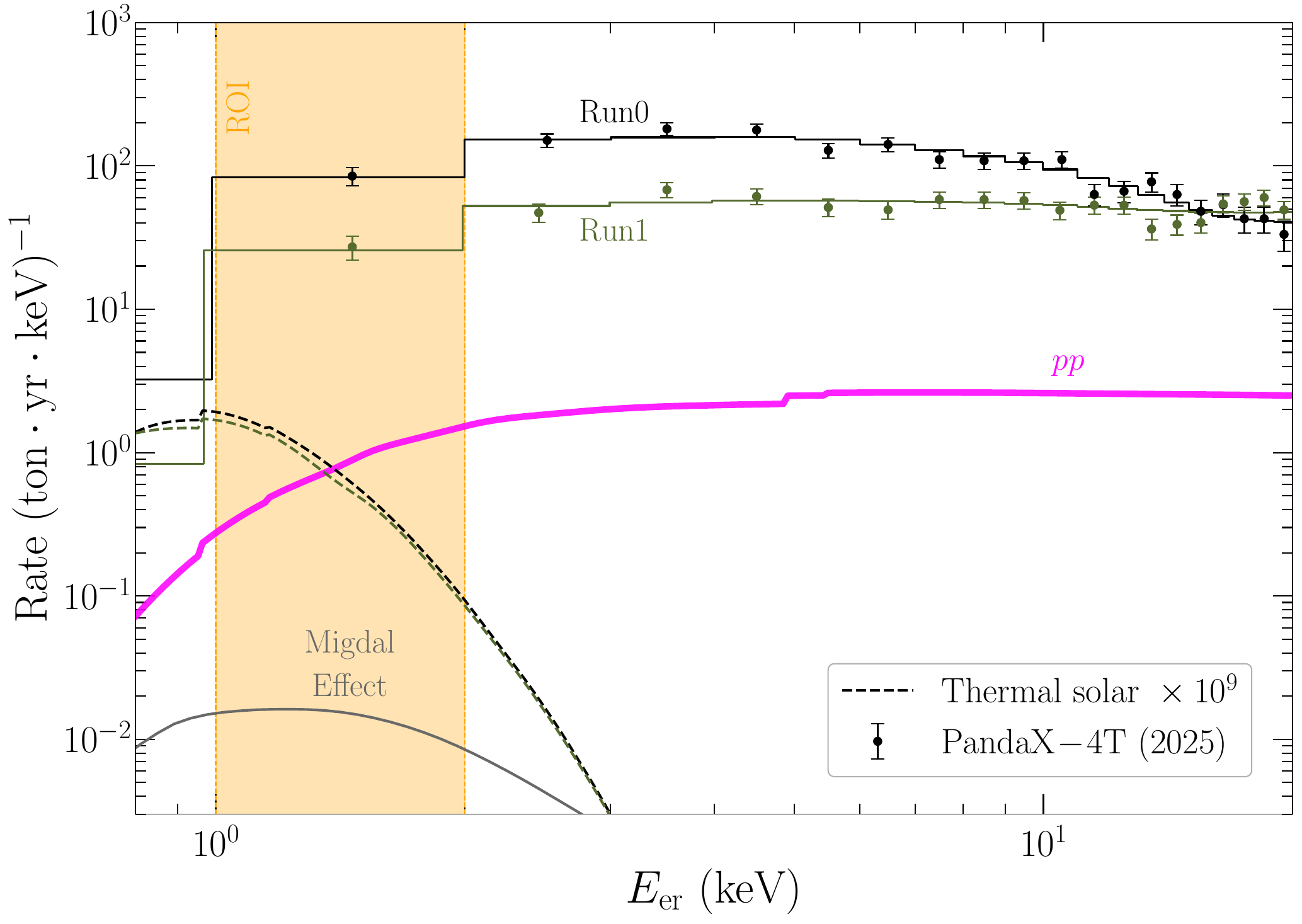}
\caption{Ionization rate of events in XENONnT, LZ and PandaX-4T (Run0 and Run1) as a function of electron energy from both $pp$ (solid magenta) and thermal solar neutrinos (dashed lines).  For thermal neutrinos, the normalization is weighted by a factor of $10^{9}$.  The data points and efficiencies (solid thick lines) from XENONnT were extracted from~\cite{XENON:2022ltv}. The data for PandaX-4T (Run0 and Run1) was taken from~\cite{PandaX:2024cic}. For LZ, we used Ref.~\cite{LZ:2025igz}. The thickness of the $pp$ neutrino rates for PandaX-4T corresponds to the combination of the two curves, each accounting for the efficiencies reported in Run0 and Run1. The solid gray lines show the Migdal effect induced electronic rate from coherent elastic neutrino-nucleus scatterings~\cite{Ibe:2017yqa,Herrera:2023xun, Blanco-Mas:2024ale,Xu:2026acq}.}
\label{rate_pandax_xenon_old_limits}
\end{figure}

In Fig.~\ref{rate_pandax_xenon_old_limits} we show the ionization event rates in XENONnT (paired) and PandaX-4T as a function of electron recoil energy, comparing the thermal neutrino signal (scaled by $\eta = 10^{9}$) with the $pp$ neutrino contribution and the experimental background models.

\begin{figure}[h]
\centering
\includegraphics[width=0.49\linewidth]{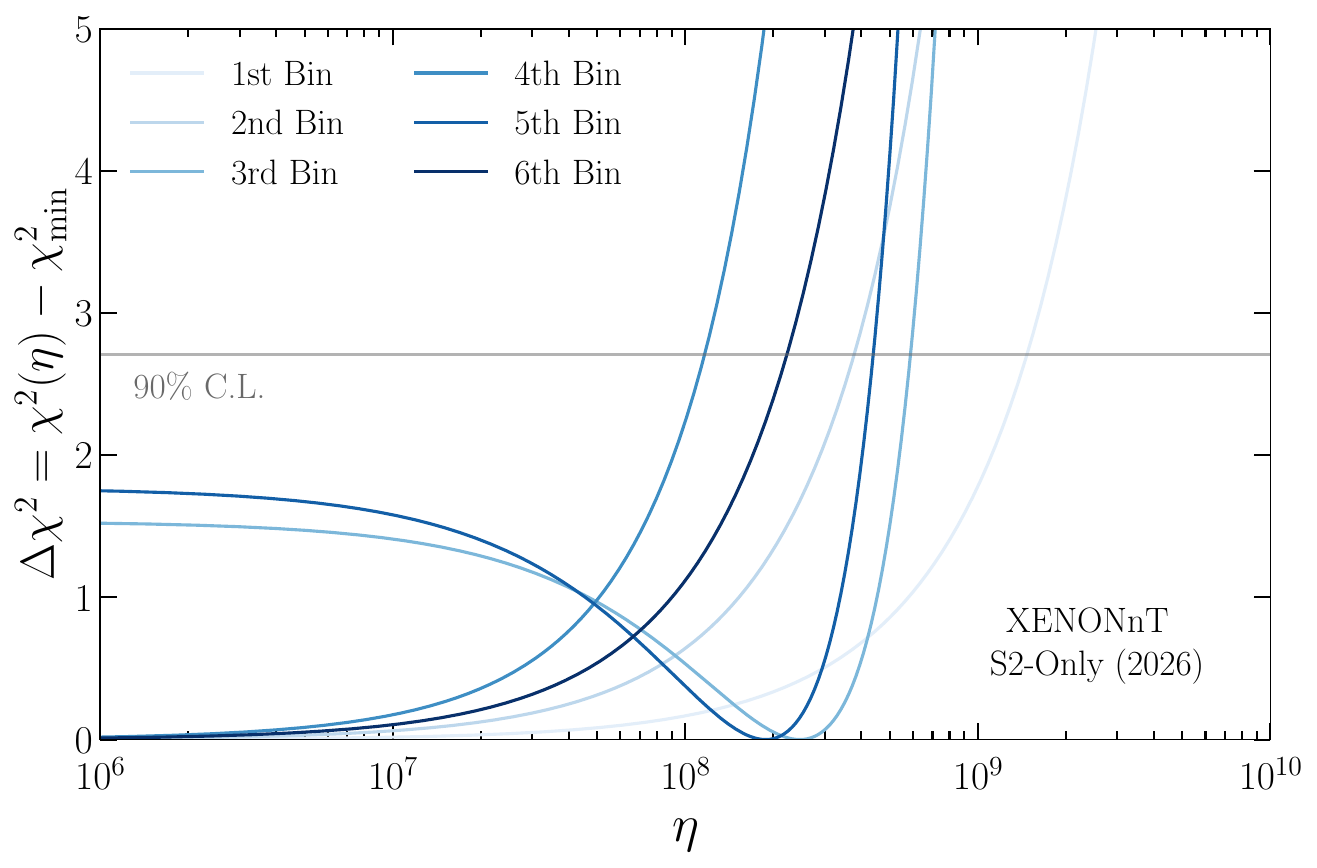}
\includegraphics[width=0.49\linewidth]{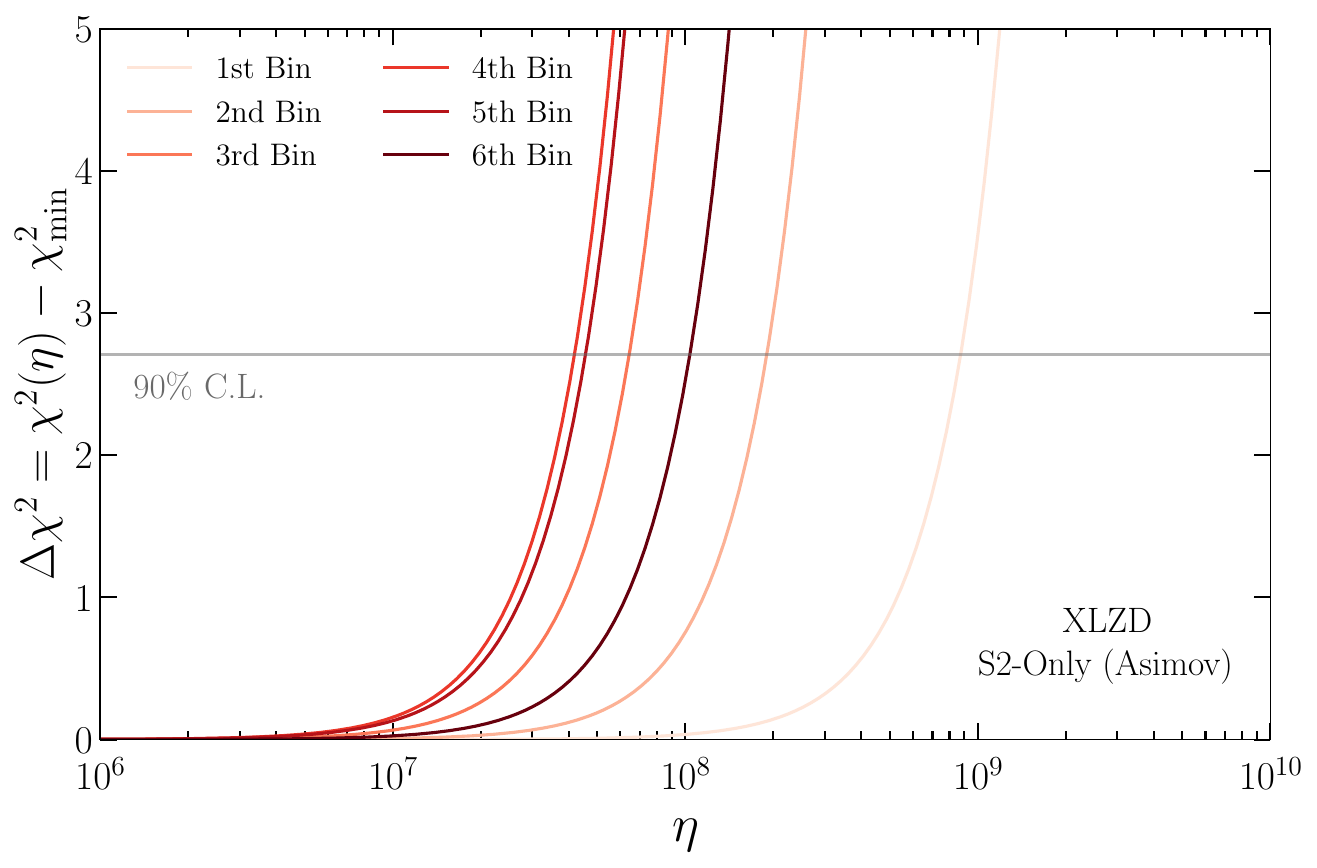}
\caption{Individual bin $\Delta \chi^{2}$ as a function of the thermal solar neutrino flux normalization $\eta$ for XENONnT (left panel) and the projected XLZD experiment (right panel).  The projection for XLZD assumes a $200$ ton-year exposure~\cite{Baudis:2024jnk}, an equal threshold and ROI as XENONnT, and a rescaling for the background and signal events as described in the main text.  The horizontal dashed line indicates the $90\%$ C.L. threshold ($\Delta \chi^2 = 2.71$).}
\label{fig:chi2_all_app}
\end{figure}

In addition to the total $\chi^2$ results shown in the main text, we have computed upper limits for each data bin shown in Fig.~\ref{fig:rate_event}. As expected, the most stringent limit comes from the 4th bin, in which the rate of neutrino-electron scattering peaks, see Table~\ref{tab:xenon_split7}. The bin-wise $\Delta \chi^2$ profiles for XENONnT and XLZD are shown in Fig.~\ref{fig:chi2_all_app}.

\begin{table}[h!]
\centering
\begin{tabular}{c S[table-format=1.2] S[table-format=1.2] S[table-format=1.2] S[table-format=1.2] S[table-format=1.2] S[table-format=1.2]}
\toprule
{Parameter} & \multicolumn{6}{c}{XENONnT S2-only 2026} \\
\cmidrule(lr){2-7}
& {1st} & {2nd} & {3rd} & {4th} & {5th} & {6th} \\
\midrule
$\eta\ (10^{9})$ & 1.47 & 0.38 & 0.59 & 0.12 & 0.44 & 0.23 \\
\bottomrule
\end{tabular}
\caption{Upper limits on the thermal neutrino flux normalization $\eta$ in units of $10^{9}$ for each bin of the XENONnT S2-only analysis~\cite{XENON:2026qow}}
\label{tab:xenon_split7}
\end{table}

\section{Cosmic-ray boosted cosmic neutrino background}
\label{sec:CRboosted}
An important question is whether other astrophysical neutrino backgrounds could contaminate the keV energy window explored by direct detection experiments. Specifically, the cosmic neutrino background (C$\nu$B), consisting of relic neutrinos that decoupled $\sim 1$~s after the Big Bang, permeates the universe with a present-day temperature $T_\nu \simeq 0.17$\,meV and a total number density $n_\nu \simeq 336~\mathrm{cm}^{-3}$ summed over all flavors~\cite{Vitagliano:2019yzm}.  The C$\nu$B has only been indirectly inferred through CMB and BBN measurements, and its direct detection remains one of the outstanding goals of astroparticle physics.

It has recently been shown that energetic cosmic rays (CRs) can scatter off the C$\nu$B via Standard Model processes, boosting relic neutrinos to high energies and producing a diffuse boosted cosmic neutrino background (DBC$\nu$B)~\cite{Ciscar-Monsalvatje:2024tvm,DeMarchi:2024zer,Herrera:2024upj,Zhang:2025rqh,Herrera:2026pzj,Cline:2026tkp}.  An analogous mechanism has been studied for supernova neutrinos~\cite{Herrera:2025pdn, Sandrock:2025nzb}.  Here we discuss whether the DBC$\nu$B could constitute a relevant background for the thermal solar neutrino search at keV energies. The DBC$\nu$B flux from CR protons scattering off relic neutrinos across all redshifts can be written as
\begin{align}
\frac{d\phi_\nu}{dT_\nu} = \int_{z_{\min}}^{z_{\max}} dz\, \frac{c}{H_0} \frac{f_i(z)\, n_\nu\,(1+z)^3}{\sqrt{\Omega_m(1+z)^3 + \Omega_\Lambda}} \int dT_p \frac{d\sigma_{p\nu}}{dT_\nu} \frac{d\phi_p}{dT_p} \,,
\label{C1}
\end{align}
where $n_\nu\simeq  336~\mathrm{cm}^{-3}$ is the present-day C$\nu$B number density, $z$ is the redshift factor, $f_i(z)$ is the CR source redshift evolution, $d\sigma_{p\nu}/dT_\nu$ is the differential proton-neutrino scattering cross section,  $d\phi_p/dT_p$ is the CR proton spectrum. The calculation includes neutral current  elastic scatterings ($\nu + p \to \nu + p$), charged current  quasi-elastic interactions, and deep inelastic scattering at high center-of-mass energies~\cite{Herrera:2026pzj}. For the cosmological parameters in Eq.~\eqref{C1}, we use the Planck 2018 values~\cite{Planck:2018vyg}: $H_0\simeq 67.4~{\rm km}\cdot{\rm s}^{-1}\cdot{\rm Mpc}^{-1}$ is the Hubble constant today, $\Omega_m\simeq 0.315$ and $\Omega_\Lambda\simeq 0.685$ are the matter and dark energy components, respectively.  

\begin{figure}[t!]
\centering
\includegraphics[width=0.65\textwidth]{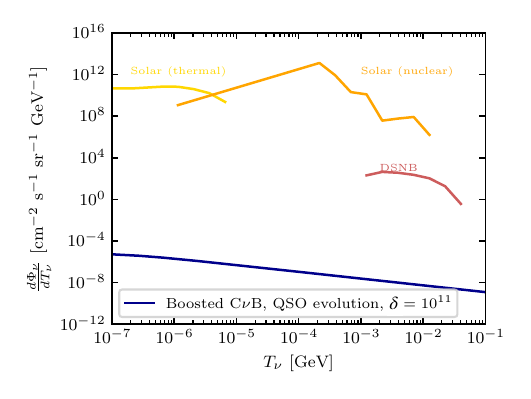}
\caption{The diffuse boosted cosmic neutrino background (DBC$\nu$B) flux arising from cosmic-ray scatterings off the C$\nu$B~\cite{Herrera:2024iye, Herrera:2026pzj}, under the quasar source evolution function, and an overdensity of $\delta=10^{11}$. The thermal solar neutrino flux (black) is shown for comparison.  The DBC$\nu$B is subdominant to the thermal solar flux by many orders of magnitude in the keV--MeV energy range relevant for direct detection experiments.}
\label{fig:DBCnuB}
\end{figure}

The resulting DBC$\nu$B flux spans a broad energy range, from $\sim$\,keV to $\sim 10^{11}$\,GeV. At the low-energy end of the DBC$\nu$B spectrum, the flux drops rapidly.  The maximum kinetic energy that a CR proton of energy $T_p$ can transfer to a relic neutrino of mass $m_\nu$ in an elastic scattering is $T_\nu^{\max} \simeq 2 T_p^2 / (T_p + 2 m_p)$ for $m_\nu \ll m_p$.  To produce a boosted neutrino with $T_\nu \sim 1$\,keV, one requires CR protons with $T_p \gtrsim 1$\,GeV.  While the CR flux is large at these energies ($d\phi_p/dT_p \sim 10^4~\mathrm{m}^{-2}\mathrm{s}^{-1}\mathrm{sr}^{-1}\mathrm{GeV}^{-1}$), the neutrino-proton cross section at $\sqrt{s} \lesssim 1$\,GeV is extremely small ($\sigma_{\nu p} \sim G_F^2 s / \pi \lesssim 10^{-45}~\mathrm{cm}^{2}$), suppressing the overall boosted flux. In particular, we find that the DBC$\nu$B flux at keV--MeV energies is several orders of magnitude below the thermal solar neutrino flux of $\Phi_{\rm th} \simeq 3.1 \times 10^6~\mathrm{cm}^{-2}\mathrm{s}^{-1}$, even when assuming an overdensity on the cosmic neutrino background of $\delta=10^{11}$ (as currently allowed by the KATRIN experiment~\cite{KATRIN:2022kkv}), see Fig.~\ref{fig:DBCnuB}. For mixed cosmic ray compositions at low energies, coherent elastic neutrino-nucleus scatterings would enhance the cross section and partially compensate this gap~\cite{Zhang:2025rqh}, but the gap remains orders of magnitude wide. The thermal solar neutrino flux therefore remains the dominant astrophysical neutrino background at keV energies by a wide margin, and the DBC$\nu$B does not constitute a relevant background for the searches discussed in this work.

\section{New Physics Contributions to the keV Neutrino Flux}
\label{app:new}
Here we consider two illustrative examples: decay of light scalar produced in the Sun, and decay of keV sterile neutrino DM in the Galaxy.  
\subsection{Light Scalar Decay} 
\label{E1}
If a light scalar particle $\phi$  couples to neutrinos via a Yukawa coupling $y_\nu$,

it can be produced inside the Sun (via plasmon decay, bremsstrahlung, neutrino pair annihilation, etc.) and can decay into neutrinos that can be an additional source of keV-scale neutrinos on Earth. However, the pure neutrino-induced production channel is always sub-dominant compared to the thermal neutrino production as it is suppressed by an extra $y_\nu^2$ in the production rate, and the current upper bounds on $y_\nu$ are of ${\cal O}(10^{-7})$ from CMB~\cite{Li:2023puz} and SN1987A~\cite{Telalovic:2024cot}. 

Therefore, we consider a scenario where the light particle production and decay are governed by different couplings. The production rate can be significantly enhanced in the solar core (compared to the neutrino case), if the particle couples to electrons or nucleons. Once produced, it decays predominantly to neutrinos, since decays to other SM final states are kinematically forbidden (except to photons, which is however loop-suppressed). Let us consider an effective Lagrangian of the type
\begin{align}
    -{\cal L} \supset y_\nu \bar{\nu}\phi \nu + y_f \bar{f}\phi f+\frac{1}{2}m_\phi^2\phi^2 \, .
\end{align}
Such interactions have been discussed in various contexts, \emph{e.g.}~scalar-portal DM~\cite{Knapen:2017xzo}, scalar NSI~\cite{Babu:2019iml} and Majorons~\cite{Heeck:2019guh}. For keV-scale $\phi$, the Yukawa couplings to electrons and nucleons are mainly constrained by stellar cooling arguments: $y_e\lesssim 10^{-16}$ and $y_N\lesssim 10^{-12}$~\cite{Knapen:2017xzo}, while the neutrino Yukawa coupling is constrained by $\Delta N_{\rm eff}$: $y_\nu\lesssim 10^{-10}$~\cite{Babu:2019iml}. Nevertheless, using either $y_e$ or $y_N$, one can still produce a large number of $\phi$ particles in the stellar core (\emph{e.g.}~via electron-ion or nucleon bremsstrahlung), which then decay to $\nu\bar{\nu}$ to give rise to an additional flux of keV neutrinos.

To determine the maximum neutrino luminosity originating from $\phi$ decay, we need to account for existing bounds based on stellar cooling arguments. For scalars produced from electron-nucleus bremsstrahlung, the energy emission rate due to scattering with a nucleus $N$ of charge $Z$ is~\cite{Yamamoto:2023zlu}
\begin{equation}
    L_{\phi} = \frac{2^3Z^2\alpha^2n_en_NT^{1/2}}{3(2\pi)^{3/2}m_N^2m_e^{3/2}}(y_Nm_e-y_em_N)^2I(m_{\phi}/T)\,,
\end{equation}
where $n_e$ is the electron density, $n_N$ is the density of nucleus $N$, $m_N$ is the mass of the nucleus, $m_e$ is the mass of the electron, $T$ is the temperature, $\alpha$ is the fine structure constant, and
\begin{align}
    I(y)=\int_y^{\infty}du\int_0^{\infty}dv\int_{-1}^1dz\int_y^{\infty}dx\,e^{-u}
    \sqrt{uv}\left(1-\frac{y^2}{x^2}\right)^{3/2}\frac{\delta(u-v-x)}{u+v-2\sqrt{uv}z}\,.
\end{align}
We sum over hydrogen, helium, nitrogen, oxygen, neon, and iron, using the abundance profiles from the solar model of Ref.~\cite{herrera_2023_10822316} (see also Refs.~\cite{Magg:2022rxb,Acharya:2024lke}).

Assuming coupling to electrons and not nucleons, we compute the luminosity of $L_\phi$ assuming the maximum electron coupling $y_e$ allowed by red giant cooling arguments~\cite{Yamamoto:2023zlu} (see also Refs.~\cite{Dev:2020eam,Balaji:2022noj,Bottaro:2023gep}) for each scalar mass.
The flux of neutrinos from $\phi\to \bar\nu\nu$ decay at the Earth, after including a Gaussian broadening of the neutrino line and the scalar decay probability, is given by 
\begin{align}
    \Phi_\nu(E_\nu)= & \frac{\dot{N}_\phi}{4\pi d^2}\frac{1}{\sqrt{2\pi\sigma^2}}\exp\left[-\frac{(E_\nu-E_0)^2}{2\sigma^2}\right]
  \left[1-\exp\left(-\frac{d}{\lambda_\phi}\right)\right] \, ,
\end{align}
where $d=1$ AU is the Earth-Sun distance, $\dot{N}_\phi=L_\phi/m_\phi$ is the number emission rate, $\lambda_\phi=\gamma c \tau_\phi$ is the decay length, with $\gamma=E_\phi/m_\phi$ being the boost factor, $\tau_\phi=1/\Gamma_\phi$ the lifetime, $\Gamma_\phi=\frac{y_\nu^2}{16\pi}m_\phi$ the decay width,  $E_0=m_\phi/2$ is the peak neutrino energy, $\sigma$ is the thermal/Doppler broadening which we take as 1.6 keV~\cite{Bahcall:1993ej}. We fix $y_\nu=10^{-10}$ (satisfying the current constraints), which implies that the decay length $\lambda_\phi\ll d$ for the entire mass range of our interest. 

Fig.~\ref{fig:majoron1} shows the maximum neutrino flux from scalar production in the Sun for different scalar masses, compared to the thermal neutrino flux and the existing and projected bounds derived in this work. While the maximum flux does not reach the level of the XENONnT sensitivity or XLZD projection, it can nonetheless be orders of magnitude larger than the thermal neutrino flux.

\subsection{keV Sterile Neutrino DM Decay}
\label{E2}

For a sterile neutrino $\nu_s$ of mass $m_s$ mixing with active neutrinos via an angle $\theta$, there are two dominant decay channels into neutrino final states: (i) $\nu_s\to \nu_a+\gamma$ (radiative) and (ii) $\nu_s\to 3\nu$ (tree-level)~\cite{Drewes:2016upu}.  The rate of process (ii) is given by
\begin{align}
\Gamma_{3\nu}=\frac{G_F^2 m_s^5}{192\pi^3}\sin^2(2\theta) \, .
\end{align}
The rate of process (i) is smaller by a factor of $27\alpha/8\pi$ and it gives a monochromatic neutrino of energy $m_s/2$. Therefore, we only consider here the 3-body decay for which the normalized neutrino spectrum per decay is given by
\begin{align}
 \frac{dN}{dE_\nu} = \frac{16}{m_s^4}E_\nu^2(3m_s-4E_\nu) \, .
\end{align}
The differential flux at Earth is given by
\begin{align}
    \frac{d\Phi}{dE_\nu} = \frac{\Gamma_{3\nu}}{4\pi m_s}\frac{dN}{dE_\nu}J_{\rm dec} \, ,
\end{align}
where $J_{\rm dec} = \int_{\rm l.o.s} \rho_s(l)dl$

is the decay $J$-factor along the line of sight (l.o.s.).  For an NFW profile~\cite{Navarro:1996gj},
\begin{equation}
    \rho_{s}(r)=\rho_s\left(\frac{r}{r_s}\right)^{-1}\left(1+\frac{r}{r_s}\right)^{-2}.
\label{eq:NFW_halo}
\end{equation}
We take $r_s=20~{\rm kpc}$ with the normalization fixed by a local DM density of $\rho_\odot=0.4~{\rm GeV\,cm^{-3}}$ at a distance from the Galactic Center $R_\odot=8.5~{\rm kpc}$.  For Milky-Way l.o.s., $J_{\rm NFW}\mathord{\sim} 10^{29}$\,eV$\cdot$cm$^{-2}$.

\begin{figure}[t!]
\includegraphics[width=0.55\textwidth]{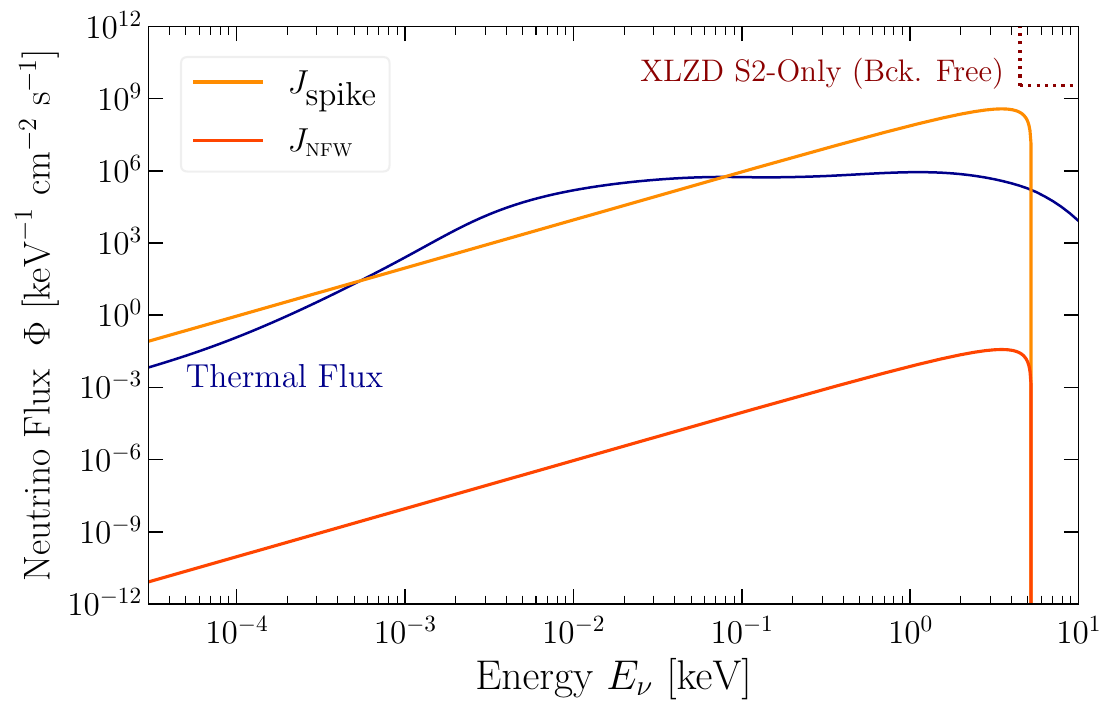}
\caption{Neutrino flux from the decay of keV sterile neutrino DM ($\nu_s \to 3\nu$) for $m_s = 7$~keV and $\sin^2 2\theta = 10^{-10}$, assuming NFW and spikes DM density profiles.  The thermal solar neutrino flux and the XLZD sensitivity are shown for comparison. 
} 
\label{fig:decay}
\end{figure}

We also consider a spike profile with~\cite{Gondolo:1999ef}
\begin{align}
    \rho_s(r) = \rho_{\rm sp}\left(1-\frac{4R_S}{r}\right)^3\left(\frac{r}{r_{\rm sp}}\right)^{-\gamma_{\rm sp}},
\end{align}
for $r_{\min}<r<r_{\rm sp}$. 
For adiabatic growth, typical spike parameters are $\gamma_{\rm sp}=7/3$, $r_{\rm sp}\mathord{\sim} 0.5$\,pc, $r_{\min}=4R_S=1.2\times 10^{-7}$\,pc for $M_{\rm BH}=4\times 10^6 M_\odot$ and $\rho_{\rm sp}=10^3$ GeV$\cdot$cm$^{-3}$. Since the annihilation rate of keV sterile neutrino DM is negligible due to the extremely small mixing angle, we use the pure (unsaturated) Gondolo-Silk spike density profile.   
In this case, the $J$-factor becomes $\mathord{\sim} 10^{39}$ eV$\cdot$ cm$^{-2}$, roughly ten orders of magnitude larger than the NFW $J$-factor. 
The resulting neutrino flux from the three-body DM decay is shown in Fig.~\ref{fig:decay} with NFW and spike profiles for a fixed DM mass of $m_s=7$ keV and $\sin^22\theta=10^{-10}$, satisfying the current X-ray constraints. 
We find that the DM-induced neutrino flux can surpass the thermal solar flux for the spike profile, indicating that future direct detection experiments could in principle probe keV sterile neutrino DM decay into neutrinos.

\end{document}